\documentclass[12pt]{article}

\usepackage{amsmath}
\usepackage{physics}
\usepackage{tensor}
\usepackage{cancel}
\usepackage{xcolor}
\usepackage{mathtools}
\usepackage[left=1in,right=1in,top=1in,bottom=1in]{geometry}
\usepackage{titling}
\usepackage{setspace}
\usepackage{dsfont}
\usepackage{amssymb}
\usepackage{cite}

\definecolor{citelink}{rgb}{0.33, 0.42, 0.18}
\definecolor{section_footnote_link}{rgb}{0.55, 0.0, 0.0}
\definecolor{equationlink}{rgb}{0.0, 0.2, 0.6}
\definecolor{urllink}{rgb}{0.53,0.15,0.34}

\usepackage[linktoc=all,colorlinks,citecolor=citelink,linkcolor=section_footnote_link,urlcolor=urllink]{hyperref}
\hypersetup{pdfauthor={C. Virally},pdftitle={Solving Superconformal Ward Identities in Mellin Space}}
\usepackage{cleveref}
\crefname{equation}{equation}{equations}
\crefname{figure}{figure}{figures}
\makeatletter
\AddToHook{cmd/appendix/before}{\def\cref@section@alias{appendix}\def\cref@subsection@alias{appendix}}
\creflabelformat{equation}{\textup{\hypersetup{linkcolor=equationlink,linkbordercolor=equationlink}(#2#1#3)}}
\makeatother

\makeatletter
\DeclareFontFamily{OMX}{MnSymbolE}{}
\DeclareSymbolFont{MnLargeSymbols}{OMX}{MnSymbolE}{m}{n}
\SetSymbolFont{MnLargeSymbols}{bold}{OMX}{MnSymbolE}{b}{n}
\DeclareFontShape{OMX}{MnSymbolE}{m}{n}{
    <-6>  MnSymbolE5
   <6-7>  MnSymbolE6
   <7-8>  MnSymbolE7
   <8-9>  MnSymbolE8
   <9-10> MnSymbolE9
  <10-12> MnSymbolE10
  <12->   MnSymbolE12
}{}
\DeclareFontShape{OMX}{MnSymbolE}{b}{n}{
    <-6>  MnSymbolE-Bold5
   <6-7>  MnSymbolE-Bold6
   <7-8>  MnSymbolE-Bold7
   <8-9>  MnSymbolE-Bold8
   <9-10> MnSymbolE-Bold9
  <10-12> MnSymbolE-Bold10
  <12->   MnSymbolE-Bold12
}{}

\let\llangle\@undefined
\let\rrangle\@undefined
\DeclareMathDelimiter{\llangle}{\mathopen}%
                     {MnLargeSymbols}{'164}{MnLargeSymbols}{'164}
\DeclareMathDelimiter{\rrangle}{\mathclose}%
                     {MnLargeSymbols}{'171}{MnLargeSymbols}{'171}
\makeatother

\numberwithin{equation}{section}

\begin{document}
    \begin{titlingpage}
        \vspace*{3em}
        \onehalfspacing
        \begin{center}
            {\LARGE Solving Superconformal Ward Identities in Mellin Space}
        \end{center}
        \singlespacing
        \vspace*{2em}
        \begin{center}
            \textbf{
            Cl\'ement Virally
            }
        \end{center}
        \vspace*{1em}
        \begin{center}
            \textsl{
            Mathematical Institute, University of Oxford, Andrew Wiles Building, Radcliffe Observatory Quarter, Woodstock Road, Oxford, OX2 6GG, U.K.
            }
            \href{mailto:clement.virally@maths.ox.ac.uk}{\small clement.virally@maths.ox.ac.uk}
        \end{center}
        \vspace*{3em}
        \begin{abstract}
            \noindent We study four-point correlators in superconformal theories in various dimensions. We develop an efficient method to solve the superconformal Ward identities in Mellin space. For 4d $\mathcal{N}=4$ SYM and the 6d $\mathcal{N}=(2,0)$ theory, our method reproduces the known solutions. As novel applications of this method, we also derive solutions in 3d $\mathcal{N} = 8$ ABJM, and in 4d $\mathcal{N} = 4$ SYM with line defects.
        \end{abstract}
    \end{titlingpage}

    \tableofcontents

    \section{Introduction}\label{sec:intro}
    The AdS/CFT correspondence~\cite{hep-th/9711200,hep-th/9802109,hep-th/9802150} is one of the most powerful tools in modern physics to study gravitational theories analytically. The most important objects and first observables to be computed using the AdS/CFT dictionary are holographic correlators. They are traditionally computed through Witten diagrams in AdS, but this technique quickly runs into technical difficulties. In recent years, additional progress has been made in many theories by bootstrapping these correlators~\cite{1608.06624,1710.05923}. Such techniques exploit principles of symmetry and impose consistency conditions on the correlators to compute them, which avoids the difficulties of the diagrammatic approach. For a review of some recent work, see~\cite{2202.08475}.

    Much recent work has been completed in Mellin space, first introduced by Mack~\cite{0907.2407}, which has been argued to be the natural language for holographic correlators~\cite{1011.1485,1107.1499,1107.1504,1111.6972,1209.4355,1410.4185}. For instance, all four-point functions have been found in supergravity with maximally superconformal backgrounds~\cite{1608.06624,1710.05923,2006.06653,2006.12505} and super Yang-Mills (SYM) in half-maximally supersymmetric backgrounds~\cite{2103.15830}. Therefore, it is desirable to understand the symmetries and consistency conditions available in Mellin space to help fix correlators.

    The simplest theories in which to study AdS/CFT are those with 16 supercharges. For boundary dimensions $d$ between 3 and 6 (above which supconformal field theories no longer exist~\cite{Nahm1978}), this means that the dual CFT can be 3d $\mathcal{N}=8$ Aharony-Bergman-Jafferis-Maldacena (ABJM) theory~\cite{0806.1218}, 4d $\mathcal{N}=4$ SYM or the 6d $\mathcal{N}=(2,0)$ theory. 

    In such superconformal theories, one of the imposed constraints are the superconformal Ward identities (SCWI). These are generally understood in position space, where they take a simple, universal form~\cite{hep-th/0405180}. However, they can also be translated into Mellin space~\cite{1608.06624,1710.05923,1712.02800}, where they take the form of difference equations. Solutions to these equations are then known in 4d $\mathcal{N}=4$ SYM~\cite{1608.06624,1710.05923} and the 6d $\mathcal{N}=(2,0)$ theory~\cite{1712.02788}. In each case, the Mellin amplitude for the four-point function of half-BPS operators $\mathcal{M}(s,t;\sigma,\tau)$ takes the form of an operator $\mathcal{D}$, built of shifts of the Mellin variables $s,t$, acting on an unconstrained ``reduced amplitude'' $\widetilde{\mathcal{M}}(s,t;\sigma,\tau)$:
    \begin{align}
        \mathcal{M}(s,t;\sigma,\tau) = \mathcal{D}\widetilde{\mathcal{M}}(s,t;\sigma,\tau)\;,\label{eq:reduced_to_full}
    \end{align}
    with simplified dependence on the R-symmetry cross-ratios $\sigma,\tau$. These operators are built out of Mellin-transforming a similar relation in position space, but so far have not been found directly in Mellin space. 

    In this paper, we develop a technique for finding such operators $\mathcal{D}$ in Mellin space, without reference to their position space representation. As applications of this new technique, we find solutions in 3d $\mathcal{N}=8$ ABJM and 4d $\mathcal{N}=4$ SYM with the insertion of line defects.

    The existence of solutions of the form of \cref{eq:reduced_to_full} in $\mathcal{N}=4$ SYM in particular has allowed for much bootstrap progress in recent years. In particular, for identical external operators with conformal dimension $\Delta=2$, the solution is such that the reduced amplitude has no dependence on $\sigma,\tau$, which leaves only a single function of $s,t$ to bootstrap. For some recent work on this theory which uses this property, see e.g. \cite{2207.01615,2412.00249,2204.07542,2209.06223,2306.12786}. Thus, finding similar results in other theories can be expected to lead to progress in the effort to bootstrap them using similar approaches.

    In maximally superconformal field theories, the half-BPS operators have protected dimensions $\Delta=\frac{d-2}{2}k$, for $k=2,3,\ldots$. For most of this paper, and unless otherwise specified, we will focus on correlators of equal operators with $k_i=2$. However, we emphasize that we expect our technique for solving the SCWI can be extended to other cases, since the 4d $\mathcal{N}=4$ solution in position space works for any $k_i$, and can be transformed into a Mellin space solution~\cite{1608.06624,1710.05923}. For correlators of equal operators, the 6d $\mathcal{N}=(2,0)$ theory position space operators also works for any $k$~\cite{1712.02788}. In \cref{sec:ki_3}, we will cover the case of $k_i=3$ in the theories in which results are known from position space. We briefly consider a mixed correlator in \cref{sec:22kk}, also in the 4d $\mathcal{N}=4$ SYM case where the answer is known.

    The rest of this paper is organized as follows. In \cref{sec:scwi}, we consider the SCWI and show how to translate them into Mellin space. In \cref{sec:pos_sol}, we review the known solutions in 4d $\mathcal{N}=4$ SYM and the 6d $\mathcal{N}=(2,0)$ theory, coming from position space. In \cref{sec:mellin_sol}, we explain how to find operators $\mathcal{D}$ which give solutions of the SCWI when acting on any function of the Mellin variables. We also show how this method reproduces the answers from position space. In \cref{sec:new_sol}, we use our approach to solve the SCWI in 3d $\mathcal{N}=8$ ABJM, which is a new result. We then review the SCWI in Mellin space for 4d $\mathcal{N}=4$ SYM with line defects and solve them, again using the same technique. We conclude in \cref{sec:conclusion}. Several longer expressions are presented in \cref{sec:full_ops,sec:higher_order_prim}, and solutions of the SCWI for other correlators can be found in \cref{sec:more_correlators}.

    \section{Superconformal Ward identities}\label{sec:scwi}
    \subsection{Position space}\label{sec:scwi_pos}
    We are interested in constructing the SCWI for the case of 4-point functions of half-BPS operators, for $3\le d\le 6$. These are superprimaries of the shortened representations of superconformal algebras. We restrict ourselves to the subset of these algebras in which the R-symmetry group is locally isomorphic to $SO(n)$. This construction was previously done in~\cite{1712.02800}, and we will follow that approach closely.

    The operators $\mathcal{O}^{I_1\ldots I_k}$ exist in the rank-$k$ symmetric traceless representation of $SO(n)$, with protected conformal dimension:
    \begin{equation}
        \Delta=\varepsilon k,\quad \varepsilon\equiv \frac{d}{2}-1\;,
    \end{equation}
    where $k=2,3,4,\ldots$ We keep track of the $R$-symmetry indices by contracting them with null vectors:
    \begin{equation}
        \mathcal{O}_k(x,t) = \mathcal{O}^{I_1\ldots I_k}(x) t_{I_1}\ldots t_{I_k},\quad t^It_I=0\;.
    \end{equation}
    The 4-point function of these operators is defined as:
    \begin{equation}
        G_{k_1k_2k_3k_4}(x_i,t_i) = \expval{\mathcal{O}_{k_1}(x_1,t_1)\mathcal{O}_{k_2}(x_2,t_2)\mathcal{O}_{k_3}(x_3,t_3)\mathcal{O}_{k_4}(x_4,t_4)}\;.
    \end{equation}
    To avoid clutter, we will often suppress the $k$-dependence of $G$ in the following. Without loss of generality, we assume $k_1\ge k_2\ge k_3\ge k_4$. Then, we can extract a kinematic factor from the 4-point function:
    \begin{equation}
        G(x_i,t_i) = \prod_{i<j}\left(\frac{t^{ij}}{x_{ij}^{2\varepsilon}}\right)^{\gamma^0_{ij}}\left(\frac{t_{12}t_{34}}{x_{12}^{2\varepsilon}x_{34}^{2\varepsilon}}\right)^L\mathcal{G}(U,V;\sigma,\tau)\;,\label{eq:G_to_scrG}
    \end{equation}
    where $L$ is known as the extremality, and $U,V,\sigma,\tau$ are cross-ratios defined below. In order to define the extremality from the indices $k$, we must distinguish two cases:
    \begin{equation}
        k_1+k_4\le k_2+k_3\ (\text{Case I}),\quad k_1+k_4>k_2+k_3\ (\text{Case II})\;.
    \end{equation}
    Then, the extremality is:
    \begin{equation}
        L = \begin{cases}
            k_4&\text{Case I}\\\frac{1}{2}(k_2+k_3+k_4-k_1)&\text{Case II}\;,
        \end{cases}
    \end{equation}
    and we have the exponents:
    \begin{align}
        \begin{split}
        \gamma_{34}^0=\gamma_{24}^0=0,\quad \gamma_{12}^0 = \tfrac{1}{2}(k_1+&k_2-k_3-k_4),\quad \gamma_{13}^0 = \tfrac{1}{2}(k_1+k_3-k_2-k_4)\\
        \gamma_{14}^0 =& \begin{cases}
            0&\text{(Case I)}\\
            \frac{1}{2}(k_1+k_4-k_2-k_3)&\text{(Case II)}
        \end{cases}\\
        \gamma_{23}^0 =& \begin{cases}
            \frac{1}{2}(k_2+k_3-k_1-k_4)&\text{(Case I)}\\
            0&\text{(Case II)}\;.
        \end{cases}
        \end{split}
    \end{align}
    The prefactors in \cref{eq:G_to_scrG} take care of the covariance under the $R$-symmetry and conformal groups, such that $\mathcal{G}$ is a function of cross-ratios invariant under these groups:
    \begin{equation}
        U = \frac{(x_{12})^2(x_{34})^2}{(x_{13})^2(x_{24})^2},\quad V = \frac{(x_{14})^2(x_{23})^2}{(x_{13})^2(x_{24})^2},\quad \sigma = \frac{(t_{13})(t_{24})}{(t_{12})(t_{34})},\quad \tau = \frac{(t_{14})(t_{23})}{(t_{12})(t_{34})}\;.
    \end{equation}
    It should be noted that $\mathcal{G}$ is a polynomial of degree $L$ in the $R$-symmetry cross-ratios $\sigma,\tau$~\cite{1710.05923}. These are the restrictions imposed by the bosonic part of the superconformal group. The fermionic generators then give us what we call the SCWI. To write them down, it is convenient to introduce new variables:
    \begin{equation}
        U=z\bar z,\quad V = (1-z)(1-\bar z),\quad \sigma = \alpha\bar\alpha,\quad \tau = (1-\alpha)(1-\bar\alpha)\;.
    \end{equation}
    In terms of these new variables, the SCWI in position space take a universal form~\cite{hep-th/0405180}:
    \begin{equation}
        \eval{(z\partial_z-\varepsilon\alpha\partial_\alpha)\mathcal{G}(z,\bar z;\alpha,\bar\alpha)}_{\alpha=1/z} = 0\;,\label{eq:scwi_pos}
    \end{equation}
    as well as an analogous equation for $\bar z$. In the next section, we will show how to translate these into Mellin space.

    \subsection{Mellin space}\label{sec:scwi_mellin}
    We will begin with a short recap of Mellin space~\cite{0907.2407} in general, and then show how to translate the SCWI into this space. In a general CFT, conformal symmetry allows us to write the 4-point function as:
    \begin{equation}
        G_{\Delta_1,\ldots,\Delta_4} (x_i) = \prod_{i<j}(x_{ij}^2)^{-\delta_{ij}^0}\mathcal{G}(\xi_r)\;,
    \end{equation}
    where the $\xi_r$ are the conformal cross-ratios. To ensure the correlation function transforms properly under conformal transformations, the exponents obey:
    \begin{equation}
        \sum_{j\neq i}\delta_{ij}^0 = \Delta_i\;.
    \end{equation}
    We then consider the connected part of the correlator, and define its Mellin transform as:
    \begin{align}
        \begin{split}
        \mathcal{G}^{\text{conn}}(x_i) = \int \frac{\dd s\dd t}{(4\pi i)^2}&U^{\frac{s}{2}-a_s}V^{\frac{t}{2}-a_t}\Gamma_{\{k_i\}}\mathcal{M}(s,t;\sigma,\tau)\\
        \Gamma_{\{k_i\}} = \textstyle{\Gamma\left(\frac{\varepsilon(k_1+k_2)-s}{2}\right)\Gamma\left(\frac{\varepsilon(k_3+k_4)-s}{2}\right)\Gamma\left(\frac{\varepsilon(k_1+k_4)-t}{2}\right)}&\textstyle{\Gamma\left(\frac{\varepsilon(k_2+k_3)-t}{2}\right)\Gamma\left(\frac{\varepsilon(k_1+k_3)-u}{2}\right)\Gamma\left(\frac{\varepsilon(k_2+k_4)-u}{2}\right)}\;,
        \end{split}\label{eq:mellin_transform}
    \end{align}
    where the integration runs parallel to the imaginary axis and we defined $a_s=\frac{\varepsilon}{2}(k_1+k_2)-\varepsilon L$, $a_t = \frac{\varepsilon}{2}\min\{k_1+k_4,k_2+k_3\}$. We also introduced the Mellin space variables $s,t,u$, which obey $s+t+u=\varepsilon\sum_{i=1}^4k_i$. Then, $\mathcal{M}(s,t;\sigma,\tau)$ is known as the Mellin amplitude.

    We now translate the SCWI, \cref{eq:scwi_pos}, into Mellin space, following an observation of \cite{1712.02800}. The differential operator $z\partial_z$ can be rewritten as:
    \begin{equation}
        z\partial_z = U\partial_U + V\partial_V - \frac{1}{1-z}V\partial_V\;.
    \end{equation}
    Then, we act on $\mathcal{G}$, but leave these derivatives as operators. We now consider the $\alpha\partial_\alpha$ piece. The $\alpha$ derivatives and evaluation at $\alpha=\frac{1}{z}$ generate rational functions. For example, we find:
    \begin{equation}
        \eval{\alpha\partial_\alpha(\sigma)}_{\alpha=1/z}=\frac{1}{z}\bar\alpha\;.
    \end{equation}
    From the fact that $\mathcal{G}$ is of degree $L$ in $\sigma,\tau$, it follows that we can multiply the result of this evaluation by $z^L(1-z)$ to find a new polynomial in $z$, of degree $z^{L+1}$. Noting that $U$ and $V$ are invariant under $z\leftrightarrow \bar z$, we can get a second copy of this relation, with $z$ instead of $\bar z$. Adding these two identities together, we find an identity which depends on $z$,$\bar z$ through functions of the form $z^n+\bar z^n$, which can easily be rewritten in terms of finite linear combinations of terms $U^mV^n$. Then, there exists a simple dictionary between the Mellin and position spaces, where $U\partial_U$, $V\partial_V$ and multiplication by $U^mV^n$ become the operators~\cite{1712.02800}:
    \begin{align}
        \begin{split}
            U\partial_U\to \left(\frac{s}{2}-\frac{\varepsilon(k_3+k_4)}{2}+\varepsilon L\right)\mathds{1}\;,\\
            V\partial_V\to \left(\frac{t}{2}-\frac{\varepsilon\min\{k_1+k_4,k_2+k_3\}}{2}\right)\mathds{1}\;,\\
            U^mV^n\to\textstyle{\left(\frac{\varepsilon(k_1+k_2)-s}{2}\right)_m\left(\frac{\varepsilon(k_3+k_4)-s}{2}\right)_m\left(\frac{\varepsilon(k_1+k_4)-t}{2}\right)_n}\times\\\textstyle{\left(\frac{\varepsilon(k_2+k_3)-t}{2}\right)_n\left(\frac{\varepsilon(k_1+k_3)-u}{2}\right)_{-m-n}\left(\frac{\varepsilon(k_2+k_4)-u}{2}\right)_{-m-n}}P_{-2m,-2n}\;,
        \end{split}\label{eq:pos_to_mellin}
    \end{align}
    where $\mathds{1}$ denotes the identity operator, $(a)_n$ is the Pochhammer symbol and $P_{m,n}$ is a shift operator, defined by:
    \begin{equation}
        P_{m,n}f(s,t) = f(s+m,t+n)\;.
    \end{equation}
    We then require that the coefficients of $\bar\alpha$ vanish independently, which gives us $L+1$ SCWI equations. In Mellin space, these are now difference relations, which is to say linear combinations of functions evaluated at different shifts, with rational function coefficients. The functions involved are the different coefficients of powers of $\sigma$ and $\tau$, which are mixed together by the SCWI.

    \section{From position to Mellin}\label{sec:pos_sol}
    Solutions to the SCWI from the previous section are known in two examples, 4d $\mathcal{N}=4$ SYM and the 6d $\mathcal{N}=(2,0)$ theory, from Mellin transforming a position space solution to the SCWI~\cite{1608.06624,1710.05923,1712.02788}. In this section, we review these solutions, of which the form will inspire the Mellin solutions we consider later.

    \subsection{4d \texorpdfstring{$\mathcal{N}=4$}{N=4} SYM solution from position space}
    For $d=4$, the SCWI are given by \cref{eq:scwi_pos} with $\varepsilon=1$:
    \begin{equation}
        \eval{(z\partial_z - \alpha\partial_\alpha)\mathcal{G}(z,\bar{z};\alpha,\bar{\alpha})}_{\alpha=1/z}=0\;.
    \end{equation}
    In position space, there is a known solution, which takes the form of a factorization of the correlator. The connected part of $\mathcal{G}$ is~\cite{1710.05923}:
    \begin{equation}
        \mathcal{G}^{\text{conn}}(z,\bar{z};\alpha,\bar\alpha) = \mathcal{G}^{\text{free,conn}}(z,\bar z;\alpha,\bar\alpha)+ R(z,\bar{z};\alpha,\bar{\alpha})\mathcal{H}(z,\bar z;\alpha,\bar\alpha)\;.
    \end{equation}
    where $\mathcal{G}^{\text{free,conn}}$ is the free part, $R$ is a known function, and $\mathcal{H}$ is the (unknown) piece which contains dynamical information. In Mellin space, it is consistent to treat the free piece as zero~\cite{1710.05923}. $R$ is given by:
    \begin{align}
        \begin{split}
        R=& (1-z\alpha)(1-\bar z\alpha)(1-z\bar\alpha)(1-\bar z\bar\alpha)\\
        =&\ \tau + (1-\sigma-\tau)V + (-\tau-\sigma\tau+\tau^2)U + (\sigma^2-\sigma-\sigma\tau)UV + \sigma V^2 + \sigma\tau U^2\;,
        \end{split}
    \end{align}
    where in the second line we rewrote $R$ in terms of the cross-ratios which can easily be translated into Mellin space using:
    \begin{align}
        \begin{split}
            U^mV^n\to\textstyle{\left(\frac{\varepsilon(k_1+k_2)-s}{2}\right)_m\left(\frac{\varepsilon(k_3+k_4)-s}{2}\right)_m\left(\frac{\varepsilon(k_1+k_4)-t}{2}\right)_n}\times\\\textstyle{\left(\frac{\varepsilon(k_2+k_3)-t}{2}\right)_n\left(\frac{\varepsilon(k_1+k_3)-u}{2}\right)_{2-m-n}\left(\frac{\varepsilon(k_2+k_4)-u}{2}\right)_{2-m-n}}P_{-2m,-2n}\;.
        \end{split}\label{eq:umvn_to_mellin_op}
    \end{align}
    This should not be confused with the similar operator in \cref{eq:pos_to_mellin}~\cite{1712.02800}. Rewriting the solution to the SCWI in Mellin space is then a matter of using these translations. We end up with a solution of the form:
    \begin{equation}
        \mathcal{M}(s,t;\sigma,\tau) = \mathcal{D}(s,t;\sigma,\tau)\widetilde{\mathcal{M}}(s,t;\sigma,\tau)\;,\label{eq:full_from_reduced}
    \end{equation}
    where $\widetilde{\mathcal{M}}(s,t;\sigma,\tau)$ is known as the reduced Mellin amplitude, and $\mathcal{D}(s,t;\sigma,\tau)$ is a difference operator obtained from the translation of $R$ into Mellin space. The reduced amplitude is obtained as a modified Mellin transform of $\mathcal{H}$:
    \begin{align}
        \begin{split}
            \mathcal{H}(x_i) = \int \frac{\dd s\dd t}{(4\pi i)^2}&U^{\frac{s}{2}-a_s}V^{\frac{t}{2}-a_t}\widetilde{\Gamma}_{\{k_i\}}\widetilde{\mathcal{M}}(s,t;\sigma,\tau)\\
            \widetilde{\Gamma}_{\{k_i\}} = \textstyle{\Gamma\left(\frac{\varepsilon(k_1+k_2)-s}{2}\right)\Gamma\left(\frac{\varepsilon(k_3+k_4)-s}{2}\right)\Gamma\left(\frac{\varepsilon(k_1+k_4)-t}{2}\right)}&\textstyle{\Gamma\left(\frac{\varepsilon(k_2+k_3)-t}{2}\right)\Gamma\left(\frac{\varepsilon(k_1+k_3)-\tilde u}{2}\right)\Gamma\left(\frac{\varepsilon(k_2+k_4)-\tilde u}{2}\right)}\;,
        \end{split}\label{eq:mellin_transform_shifted}
    \end{align}
    where we introduced $\tilde u\equiv u-4$. As we'll discuss later, this definition makes crossing symmetry manifest in the reduced amplitude.

    The operator $\mathcal{D}$ is what we call the solution to the SCWI. We present it explicitly for this theory in \cref{eq:n4_d}. It can act on any function $f(s,t)$ to produce a solution to the Ward identities. Thus, once this operator is known, it suffices to work with reduced amplitudes $\widetilde{\mathcal{M}}(s,t)$, which are much simpler than the full amplitude. For example, for the case of $k_i=2$ which we will consider in most of this paper, the reduced amplitude is a single function independent of $\sigma,\tau$, instead of 6 functions (as the different coefficients of $\sigma,\tau$) connected by the SCWI.

    One important constraint to impose beyond the SCWI is crossing symmetry. Note that, since $\mathcal{D}$ can make any function of $s,t$ a solution to the SCWI, it does not impose crossing automatically. Instead, crossing symmetry is imposed on the reduced correlators, in a modified form. When the conformal dimensions are all equal ($k_i=L$), crossing symmetry in the case of the usual Mellin amplitude reads:
    \begin{equation}
        \mathcal{M}(s,t;\sigma,\tau) = \sigma^L\mathcal{M}(u,t;1/\sigma,\tau/\sigma) = \tau^L\mathcal{M}(t,s;\sigma/\tau,1/\tau)\;,\label{eq:crossing}
    \end{equation}
    where $s+t+u=4\varepsilon L$. However, in the reduced Mellin amplitude, crossing symmetry becomes instead~\cite{1710.05923}:
    \begin{equation}
        \widetilde{\mathcal{M}}(s,t;\sigma,\tau) = \sigma^{L-2}\widetilde{\mathcal{M}}(\tilde u,t;1/\sigma,\tau/\sigma) = \tau^{L-2}\widetilde{\mathcal{M}}(t,s;\sigma/\tau,\tau/\sigma)\;,\label{eq:crossing_reduced}
    \end{equation}
    where $\tilde u = u-4$. This generalizes to the case of unequal conformal dimensions. In other theories, the structure will be much the same, although the shift in $u$ will be different. Thus, in searching for an operator $\mathcal{D}$, we will ignore additional constraints, including crossing symmetry, and adopt the philosophy that we should instead impose these as constraints on the reduced Mellin amplitude. We will discuss how to do this in each theory in its own section.

    \subsection{6d \texorpdfstring{$\mathcal{N}=(2,0)$}{N=(2,0)} theory solution from position space}
    We now briefly present the solution of the SCWI in the 6d $\mathcal{N}=(2,0)$ theory, as found in~\cite{1712.02788}. This operator is valid for any correlator of four identical operators of conformal dimension $2k$, which we leave generic for now. In position space, the connected part of the correlator is expressed generically as:
    \begin{equation}
        \mathcal{G}^{\text{conn}}(U,V;\sigma,\tau) = \mathcal{G}^{\text{free,conn}}(U,V;\sigma,\tau)+\Upsilon \circ \mathcal{H}_k(U,V;\sigma,\tau)\;,
    \end{equation}
    with $\Upsilon$ a somewhat complicated differential operator, which interested readers can find in~\cite{1712.02788}, and $\mathcal{H}_k(U,V;\sigma,\tau)$ an unconstrained function, which we call the reduced amplitude. As before, the Mellin transform of the free part can be treated as zero~\cite{1710.05923}. This expression can be Mellin transformed, as we did in 4d $\mathcal{N}=4$ SYM, to give the action of an operator $\mathcal{D}$ on a reduced amplitude, as in \cref{eq:full_from_reduced}. In order to write $\mathcal{D}$, we first have to write the reduced amplitude as:
    \begin{equation}
        \widetilde{\mathcal{M}}(s,t;\sigma,\tau) = \sum_{l+m+n=k-2}\sigma^m\tau^n \widetilde{\mathcal{M}}_{k,lmn}(s,t)\;.\label{eq:6d_red_amp_sep}
    \end{equation}
    For $k=2$, which we will focus on in most of the text, this separation is not needed (we only get the $l=m=n=0$ component), but this will be important in \cref{sec:ki_3}, where we treat the $k_i=3$ case. The operator which solves the SCWI, obtained from position space, is written as:
    \begin{equation}
        \mathcal{D}=-\frac{1}{4a^2}\big((XY)\widehat{B\mathfrak{R}}+(XZ)\widehat{C\mathfrak{R}}+(YZ)\widehat{A\mathfrak{R}}\big)\;,
    \end{equation}
    where:
    \begin{equation}
        \mathfrak{R}=a^2 B C+ b^2AC+ c^2A B+a b C (-A-B+C)+ac B (-A+B-C)+b c A (A-B-C)\;,
    \end{equation}
    and $X,Y,Z$ are shifted versions of the Mellin variables that depend on $l,m,n$:
    \begin{equation}
        X = s+4l-4k+2,\quad Y = t+4n-4k+2,\quad Z = u+4m-4k+2\;.\label{eq:6d_xyz}
    \end{equation}
    To understand this expression, one needs to first expand it into terms of the form $A^\alpha B^{3-\alpha-\beta}C^\beta$, which we can then interpret as shift operators acting to the right of the $X,Y,Z$ terms:
    \begin{equation}
        A^\alpha B^{3-\alpha-\beta}C^\beta \to \frac{\Gamma^2(-\frac{s}{2}+2k+\alpha)\Gamma^2(-\frac{t}{2}+2k+\beta)\Gamma^2(-\frac{u}{2}+2k+(3-\alpha-\beta))}{\Gamma^2(-\frac{s}{2}+2k)\Gamma^2(-\frac{t}{2}+2k)\Gamma^2(-\frac{u}{2}+2k)}P_{-2\alpha,-2\beta}\;.
    \end{equation}
    The R-symmetry factors are encoded in $a,b,c$:
    \begin{equation}
        \sigma = \frac{b}{a},\quad \tau = \frac{c}{a}\;.
    \end{equation}
    Crossing symmetry is once again expressed in terms of a shifted $u$ variable in this theory. As before, the full amplitude is required to obey \cref{eq:crossing}, which requires that the reduced amplitude obey \cref{eq:crossing_reduced} with $\tilde u=u-6$.

    The full form of the operator will be presented in \cref{sec:n20d6} and \cref{sec:6d_full_op}, when we solve for it directly in Mellin space. For now, the important lesson of 6d $\mathcal{N}=(2,0)$ is that, once again, the solution of the SCWI in Mellin space can be written as a linear combination of shift operators acting on some arbitrary function, with the coefficients of the shifts given by rational functions. We will now take this as an assumption and show how we can find such an operator without knowledge of the position space solution.

    \section{Solving the SCWI in Mellin space}\label{sec:mellin_sol}
    In this section, we explain how to derive solutions to the SCWI directly in Mellin space. The fundamental idea is that we use the fact that the SCWI are given by the action of a difference operator, i.e. they are given by $L+1$ equations of the form:
    \begin{equation}
        \mathcal{W}_i f(s,t;\sigma,\tau)=0\;,\label{eq:scwi_mellin_op}
    \end{equation}
    where:
    \begin{equation}
        \mathcal{W}_i = \sum_{m,n} \ell^i_{m,n}(s,t;\sigma,\tau) P_{m,n}\;,
    \end{equation}
    and the $\ell^i_{m,n}$ are rational function coefficients. Following the examples of the previous section, we assume that the solution to such equations is given by a new difference operator, $\mathcal{D}$, acting on an arbitrary function $f(s,t)$ such that:
    \begin{equation}
        \mathcal{W}_i\mathcal{D} \widetilde{f}(s,t;\sigma,\tau)=0\label{eq:D_def}
    \end{equation}
    for \emph{any} arbitrary function $\widetilde{f}(s,t;\sigma,\tau)$. Our objective is thus to reconstruct this $\mathcal{D}$ operator from the Ward identities themselves. We assume that the SCWI are solved by an operator built from linear combinations of shifts:
    \begin{equation}
        \mathcal{D} = \sum_{m,n}\sum_{i+j\le2} h^{i,j}_{m,n}(s,t)\sigma^i\tau^j P_{m,n}\;.\label{eq:D_ansatz}
    \end{equation}
    Thus, constructing the operator means understanding both the $P_{m,n}$ content in the operator, and the coefficients $h^{i,j}_{m,n}$. To do this, we move on to the central technical observation of this paper. We consider the action of a shift operator $P_{m,n}$ on a function of the form $X^sY^t$, for arbitrary auxiliary variables $X,Y$:
    \begin{equation}
        P_{m,n}X^sY^t = (X^mY^n)(X^{s}Y^{t})\;.
    \end{equation}
    The prefactor, the eigenvalue of the $P_{m,n}$ operator corresponding to this function, is simply built of powers of $X,Y$. But then, this means that, if we act with $\mathcal{D}$ on this function, we get a sum of terms in which the powers of $X,Y$ tell us exactly which shift operator created each term. Of course, by definition of $\mathcal{D}$, such an expression solves the SCWI.

    In general, if an operator $\mathcal{D}$ of the form we expect exists, there is therefore a solution to the SCWI of the form $\mathcal{D} X^s Y^t$. Since different shifts are separated by this function, plugging the ansatz for $\mathcal{D}$ in \cref{eq:D_ansatz} into \cref{eq:D_def} and requiring it to be solved for all $X,Y$ gives us a way to find the $h_{m,n}^{i,j}(s,t)$. In practice, however, it is preferable to first find the shift operators that show up, and then solve for the coefficients.

    Since the coefficients of the shift operators are rational functions in the known examples, we assume they will be as well in the other theories. This is also justified by the fact that the SCWI have rational coefficients for their shift operators. Then, by multiplying $X^sY^t$ by a sufficient polynomial before acting with $\mathcal{D}$, we can create a solution to the SCWI which is a polynomial times $X^sY^t$.

    We can now invert this logic to find the individual shifts without knowledge of the full operator. Assuming $\mathcal{D}$ exists, we search for a solution to the SCWI of the form:
    \begin{equation}
        f(s,t;\sigma,\tau) = X^sY^t Q_{X,Y}(s,t;\sigma,\tau)\;,\label{eq:q_from_f}
    \end{equation}
    where $Q_{X,Y}$ is a polynomial in $s,t,\sigma,\tau$, of which the coefficients depend on powers of $X,Y$. This can be done simply by assuming a polynomial ansatz and putting it in the SCWI equations. Then, assuming such a polynomial comes from the action of $\mathcal{D}$ on some primitive function, the powers of $X,Y$ that appear tell us the shift operator content of $\mathcal{D}$. 

    Thus, once we find such a solution, we can read off which $m,n,i,j$ to sum in \cref{eq:D_ansatz}, by considering which powers of $X$ and $Y$ appear. We then act on $X^sY^t$ by an operator of the form of \cref{eq:D_ansatz}, and demand that this solves the SCWI for any $X,Y$. This generates equations which relate the $h^{i,j}_{m,n}(s,t)$ functions to each other, which we can solve up to ambiguities, discussed in the next section. In \cref{sec:n4d4}, we will go through the example of solving the SCWI in 4d $\mathcal{N}=4$ SYM in some detail in order to fully illustrate the method of this section.

    We now discuss how to obtain the crossing constraints for the reduced amplitudes. Based on the examples of the previous sections, crossing symmetry remains largely unchanged in the reduced amplitude, except for the fact that $u$ gets shifted. Given this, we can directly evaluate \cref{eq:crossing}, assuming that each of these amplitudes comes from a different primitive, i.e.:
    \begin{align}
        \begin{split}
        \mathcal{M}(s,t;\sigma,\tau) = \mathcal{D} f_1(s,t;\sigma,\tau),&\quad \sigma^L\mathcal{M}(u,t;\tfrac{1}{\sigma},\tfrac{\tau}{\sigma}) = \mathcal{D}f_2(s,t;
        \sigma,\tau),\\
        \tau^{L}\mathcal{M}(t,s;\sigma,\tau) &= \mathcal{D}f_3(s,t;\sigma,\tau)\;.
        \end{split}
    \end{align}
    Then, we will find that, for our choices of the ambiguities described below, $f_1,f_2,f_3$ are related by the reduced crossing equations, \cref{eq:crossing_reduced}, for some particular choice of a constant $A$ such that $s+t+\tilde u=A$, which tells us the shift in $u$.

    \subsection{Ambiguities of the Mellin solution}\label{sec:ambiguities}
    There is an inherent ambiguity to the solution in Mellin space, which appears as a redefinition of the operator $\mathcal{D}$ by an arbitrary right-multiplication. For any function $h(s,t)$, we could have defined an operator $\mathcal{D}_h$ by:
    \begin{equation}
        \mathcal{D}_hf(s,t) = \mathcal{D} h(s,t)f(s,t)\;.
    \end{equation}
    Of course, since $h(s,t)f(s,t)$ is a function of $s,t$, this still solves the SCWI by definition of $\mathcal{D}$. The ambiguity is, in fact, not limited to right-multiplication by functions, but can involve the right-action of any operator. We could, for example, take fixed  functions $g,h$ of $s,t$ and any numbers $m,n$, and then define:
    \begin{equation}
        \mathcal{D}_{g,h}f(s,t) = \mathcal{D} (g(s,t)f(s,t)+ h(s,t)P_{m,n}f(s,t))\;,
    \end{equation}
    or add a more complicated set of shifts in this manner. This emerges as higher-order polynomials of $X,Y$ in the solution for $Q_{X,Y}$, as well as a set of arbitrary leftover functions after fixing the $h^{i,j}_{m,n}(s,t)$ coefficients. To fix the ambiguities in the number of shift operators, we choose to take the minimal number of shifts, equivalent to choosing a $Q_{X,Y}(s,t)$ with the lowest order in $X,Y$ that solves the SCWI. We can then exploit the leftover ambiguity coming from the choice of a function to give $\mathcal{D}$ operators desirable properties, such as taking crossing-symmetric polynomials to crossing-symmetric polynomials. We will often make this particular choice in the following sections.

    \subsection{4d \texorpdfstring{$\mathcal{N}=4$}{N=4} SYM}\label{sec:n4d4}
    Following the method of this section, we can solve the SCWI for 4d $\mathcal{N}=4$ SYM. To begin, we want to find the shift operators that actually appear. For this, we specialize to the case where $k_i=2$ and write down an ansatz of the form:
    \begin{equation}
        Q_{X,Y}(s,t;\sigma,\tau) = \sum_{i=0}^2\sum_{j=0}^{2-i}\sum_{a,b}\sigma^i\tau^j X^aY^b Q^{i,j,a,b}(s,t)\;,
    \end{equation}
    where the sum over $i,j$ is fixed such that this is a polynomial of order 2 in $\sigma,\tau$, but the sum over $a,b$ is unfixed at this stage. We want to find a solution to the Ward identities of the form of \cref{eq:q_from_f}, with $Q_{X,Y}$ as above. We find that $Q^{i,j,a,b}_{X,Y}$ should be of order 4 in $s,t$. 

    We choose to have shifts in the negative $s,t$ directions, corresponding to negative powers of $X,Y$, with the lowest possible number of shifts allowed being 0. This is to match both the shifts in the SCWI, and the known position space result. Of course, this is simply equivalent to a choice of an overall shift in the Mellin space ambiguity of our answer. With this choice, and with the minimal orders of $X,Y$ possible, we find a solution of the SCWI of the form:
    \begin{align}
        \begin{split}
            f(s,t)={}&(t-4)^2(s+t-4)^2X^sY^{t-2}+\tau^2(s-4)^2 (s+t-4)^2 X^{s-2} Y^t+\sigma^2(s-4)^2 (t-4)^2 X^{s-2} Y^{t-2}\\ 
            +&\tau\left(-(s-4)^2 (s+t-4)^2 X^{s-2} Y^t-(t-4)^2
            (s+t-4)^2 X^s Y^{t-2}+(s+t-4)^2
            (s+t-2)^2 X^s Y^t\right)\\
            +&\sigma\left( X^sY^{t-4}(t-6)^2(t-4)^2 - \sigma (t-4)^2(s-4)^2 X^{s-2}Y^{t-2} - \sigma(t-4)^2(s+t-4)^2X^sY^{t-2}\right)\\
            +&\sigma\tau\left((s-6)^2 (s-4)^2 X^{s-4}Y^t-(s-4)^2 (t-4)^2 X^{s-2}Y^{t-2}-(s-4)^2 (s+t-4)^2 X^{s-2}Y^t\right)\;.
        \end{split}
    \end{align}
    From this, we can read off the shifts by counting powers of $X,Y$:
    \begin{align}
        \begin{split}
        \mathcal{D} ={}& h^{0,0}_{0,-2}(s,t)P_{0,-2}+\tau^2 h^{0,2}_{-2,0}(s,t)P_{-2,0} + \sigma^2h^{2,0}_{-2,-2}(s,t)P_{-2,-2}\\
        &+\tau\left(h^{0,1}_{-2,0}(s,t)P_{-2,0}+h^{0,1}_{0,-2}(s,t)P_{0,-2}+h^{0,1}_{0,0}(s,t)P_{0,0}\right)\\
        &+\sigma\left(h^{1,0}_{0,-4}(s,t)P_{0,-4}+h^{1,0}_{-2,-2}(s,t)P_{-2,-2}+h^{1,0}_{0,-2}(s,t)P_{0,-2}\right)\\
        &+\sigma\tau\left(h^{1,1}_{-4,0}(s,t)P_{-4,0}+h^{1,1}_{-2,-2}(s,t)P_{-2,-2}+h^{1,1}_{-2,0}(s,t)P_{-2,0}\right)\;,\label{eq:n4_scwi_xy_sol}
        \end{split}
    \end{align}
    for some, as of now undetermined, functions $h^{i,j}_{m,n}(s,t)$. In order to determine these functions, we act with $\mathcal{D}$, in the above form, on $X^sY^t$ once more, and demand that the resulting quantity solve the SCWI for any $X,Y$. This gives us equations which fix the $h^{i,j}_{m,n}(s,t)$ as functions of each other. Thus we obtain our final result, up to the Mellin ambiguity which appears as a right-multiplication by one of the $h^{i,j}_{m,n}(s,t)$, left unfixed by our method. To write the operator explicitly, we separate it into components:
    \begin{equation}
        \mathcal{D} = \sum_{i=0}^2\sum_{j=0}^{2-i}\sigma^i\tau^j \mathcal{D}_{i,j}\;.\label{eq:D_separation}
    \end{equation}
    Then, we have the 4d $\mathcal{N}=4$ SYM answer:
    \begin{align}
        \begin{split}
            \mathcal{D}_{0,0} ={}& (t-4)^2(s+t-4)^2 P_{0,-2}\; ,\\
            \mathcal{D}_{0,1} ={}& -(s-4)^2 (s+t-4)^2P_{-2,0}-(t-4)^2
            (s+t-4)^2P_{0,-2}+(s+t-2)^2 (s+t-4)^2P_{0,0}\;,\\
            \mathcal{D}_{0,2} ={}& (s-4)^2 (s+t-4)^2P_{-2,0}\;,\\
            \mathcal{D}_{1,0} ={}& -(s-4)^2 (t-4)^2 P_{-2,-2}+(t-6)^2(t-4)^2 P_{0,-4}-(t-4)^2 (s+t-4)^2
            P_{0,-2}\; ,\\
            \mathcal{D}_{1,1} ={}& (s-6)^2 (s-4)^2P_{-4,0}-(s-4)^2 (t-4)^2P_{-2,-2}-(s-4)^2 (s+t-4)^2 P_{-2,0}\;,\\
            \mathcal{D}_{2,0} ={}& (s-4)^2(t-4)^2P_{-2,-2}\;,\label{eq:n4_d}
        \end{split}
    \end{align}
    where we chose the Mellin ambiguity to be the minimal-order one to take polynomials to polynomials, and to be compatible with (modified) crossing symmetry, \cref{eq:crossing_reduced} with $\tilde u = u-4$. It also matches the operator obtained as a solution from position space, up to a numerical factor. We may notice that the polynomial coefficients above are the same as the ones in \cref{eq:n4_scwi_xy_sol}. This is no coincidence, since we imposed in deriving that equation that we should find a minimal-order polynomial solution, and in deriving the general form we studied the action of $\mathcal{D}$ on $X^sY^t$, requiring that the coefficients be polynomials using the ambiguity.

    Therefore, the primitive for the SCWI solution in \cref{eq:n4_scwi_xy_sol} under the action of this operator is exactly $X^sY^t$, with no polynomial prefactor. This will only occur when we take the minimal polynomials in each step of the computation of the operators, and impose the right conditions on the Mellin ambiguity.

    Our conventions for the Mellin ambiguity of the operator $\mathcal{D}$ are also the ones which reproduce the primitive for the supergravity amplitude found in~\cite{1710.05923}, as we will now see. The supergravity amplitude can be found in Mellin space in~\cite{2006.12505}, and reproduces a result of~\cite{hep-th/0002170}. Up to a normalization factor, it is given by:
    \begin{align}
        \begin{split}
        &\mathcal{M}_{\text{sugra}}(s,t;\sigma,\tau) = \mathcal{M}_{\text{sugra},s}(s,t;\sigma,\tau)+\mathcal{M}_{\text{sugra},t}(s,t;\sigma,\tau)+\mathcal{M}_{\text{sugra},u}(s,t;\sigma,\tau)\;,\\
        &\mathcal{M}_{\text{sugra},s}(s,t;\sigma,\tau) = -\frac{(t-4)(u-4)+(s+2)((t-4)\sigma+(u-4)\tau)}{s-2}\;,\\
        &\mathcal{M}_{\text{sugra},t}(s,t;\sigma,\tau) = \tau^2\mathcal{M}_{\text{sugra},s}\left(t,s,\tfrac{\sigma}{\tau},\tfrac{1}{\tau}\right)\,,\quad \mathcal{M}_{\text{sugra},u}(s,t,\sigma,\tau)=\mathcal{M}_{\text{sugra},s}\left(u,t;\tfrac{1}{\sigma},\tfrac{\tau}{\sigma}\right)\;,
        \end{split}
    \end{align}
    which has the primitive:
    \begin{equation}
        \widetilde{\mathcal{M}}_{\text{sugra}} = \frac{1}{(s-2)(t-2)(\tilde u-2)}\;.
    \end{equation}
    We can explicitly see that this last amplitude is crossing symmetric with respect to $s,t,\tilde u$.
    
    \subsection{6d \texorpdfstring{$\mathcal{N}=(2,0)$}{N=(2,0)} theory}\label{sec:n20d6}
    We now turn to the case of 6d $\mathcal{N}=(2,0)$ theory, and solve the SCWI in Mellin space. We apply the techniques of this section and decompose the resulting operator as in \cref{eq:D_separation}. Then, we can write the $\mathcal{D}_{0,0}$ component of the solution as:
    \begin{align}
        \begin{split}
            \mathcal{D}_{0,0}={}&-(t-10) (t-8)^2 (s+t-10) (s+t-8)P_{0,-4}\\
            &-(s-8)(t-6)(t-8)(s+t-10)(s+t-8)P_{-2,-2}\\
            &+(t-6) (t-8) (s+t-8)^2 (s+t-6)P_{0,-2}\;.
        \end{split}\label{eq:6d_d_00}
    \end{align}
    Due to its length, we present the rest of the operator in \cref{sec:6d_full_op}. We also find that crossing symmetry is given by \cref{eq:crossing_reduced}, with $\tilde u=u-6$ as we expected from the position space solution for this theory.

    We are now interested in considering the supergravity amplitude, which, up to a normalization factor, is~\cite{2006.12505}:
    \begin{align}
        \begin{split}
            \mathcal{M}_{\text{sugra}}(s,t;\sigma,\tau)={}&\mathcal{M}_{\text{sugra},s}(s,t;\sigma,\tau)+\mathcal{M}_{\text{sugra},t}(s,t;\sigma,\tau)+\mathcal{M}_{\text{sugra},u}(s,t;\sigma,\tau)\;,\\
            \mathcal{M}_{\text{sugra},s}(s,t;\sigma,\tau)={}&-\bigg(\frac{(t-8)(u-8)+(s+2)((t-8)\sigma+(u-8)\tau)}{s-4}\\
            &+\frac{(t-8)(u-8)+(s+2)((t-8)\sigma+(u-8)\tau)}{4(s-6)}\bigg)\;,\\
            \mathcal{M}_{\text{sugra},t}(s,t;\sigma,\tau)={}& \tau^2\mathcal{M}_{\text{sugra},s}(t,s;\tfrac{\sigma}{\tau},\tfrac{1}{\tau})\,,\quad \mathcal{M}_{\text{sugra},u}(s,t;\sigma,\tau)= \sigma^2\mathcal{M}_{\text{sugra},s}(u,t;\tfrac{1}{\sigma},\tfrac{\tau}{\sigma})\;.
        \end{split}
    \end{align}
    The primitive for this amplitude is simply:
    \begin{equation}
        \widetilde{\mathcal{M}}_{\text{sugra}}(s,t) = -\frac{1}{8(s-4)(t-4)(\tilde u-4)}\;.
    \end{equation}
    In \cite{1712.02788}, a different primitive was found for this amplitude:
    \begin{equation}
        \overline{\mathcal{M}}_{\text{sugra}}(s,t)\propto\frac{1}{(s-6)(s-4)(t-6)(t-4)(\tilde u-4)(\tilde u-6)}\;.
    \end{equation}
    The difference between these solutions comes from the lack of a factor of
    \begin{equation}
        f_{\text{amb}}=(s-6)(t-6)(\tilde u-6)
    \end{equation}
    in our Mellin ambiguity compared to the one picked out by the position space representation. We choose to keep the ambiguity choice that led to our version of $\mathcal{D}$ as it allows for lower-order polynomial solutions to the SCWI to have polynomial primitives. Indeed, the lowest-order crossing-symmetric polynomial solution is the order-4 solution:
    \begin{align}
        \begin{split}
            \mathcal{M}^{(4)}(s,t;\sigma,\tau) ={}& 8 (t-8) (s+t-8) (s (2 t-13)+2 (t-16) t+124)\\
            &+8 \tau(s+t-8) \left(s^2 (4 t+3)+2 s (2 (t-8) t-75)+3
            (t-50) t+1168\right)\\
            &+8 \tau^2(s-8) (s+t-8) (2 s (s+t-16)-13 t+124)\\
            &-8 \sigma(-8 + t) (-464 + 3 t (18 + t) + 2 s (-16 + t) (-13 + 2 t) + s^2 (-26 + 4 t))\\
            &-8\sigma\tau (-8 + s) (-464 - 26 (-16 + t) t + s^2 (3 + 4 t) + s (54 - 90 t + 4 t^2))\\
            &+8\sigma^2 (-8 + s) (-8 + t) (84 - 13 t + s (-13 + 2 t))\;,
        \end{split}
    \end{align}
    which has been normalized to have the primitive:
    \begin{equation}
        \widetilde{\mathcal{M}}^{(4)}(s,t) = 1
    \end{equation}
    under the action of $\mathcal{D}$, but would have a non-trivial primitive:
    \begin{equation}
        \overline{\mathcal{M}}^{(4)}(s,t) \propto \frac{1}{(s-6)(t-6)(\tilde u-6)}
    \end{equation}
    under the operator obtained from position space.

    \section{New solutions in Mellin space}\label{sec:new_sol}
    \subsection{3d \texorpdfstring{$\mathcal{N}=8$}{N=8} ABJM}\label{sec:n8d3}
    In 3d $\mathcal{N}=8$ ABJM, we are not aware of a known solution to the SCWI in Mellin space from the position representation. We apply the techniques of \cref{sec:mellin_sol}, and separate the operator as in \cref{eq:D_separation}. The $\mathcal{D}_{0,0}$ part, for example, is then:
    \begin{align}
        \begin{split}
            \mathcal{D}_{0,0} ={}& (t-6) (t-4) (t-2)^2 (s+t-2) (s+t-1)P_{0,-6}\\
            &+(s-4) (s-2) (t-3) (t-2)(s+t-2) (s+t-1) P_{-4,-2}\\
            &+(s-2)
            (t-4) (2 t-5) (t-2) (s+t-2) (s+t-1)
            P_{-2,-4}\\
            &-(s-2) (t-3) (t-2) (s+t-2)
            (s+t) (2 s+2 t-3)P_{-2,-2}\\
            &-(t-4)(t-2) (s+t-2) (s+t) (s (2 t-5)+2
            (t-4) t+5)P_{0,-4}\\
            &+(t-3) (t-2)
            (s+t-2)^2 (s+t) (s+t+2)P_{0,-2}\;.\label{eq:3d_d_00}
        \end{split}
    \end{align}
    Due to the length of the expressions involved, the rest of the operator can be found in \cref{sec:3d_full_op}. In this theory, we find that crossing-symmetric reduced amplitudes are, as before, formulated in terms of $s,t,\tilde u$, which obey:
    \begin{equation}
        s+t+\tilde u = -4\;.\label{eq:stu_3d}
    \end{equation}
    Since $s+t+u=4$, this gives us $\tilde u\equiv u-8$. Then, crossing symmetry for the reduced amplitude is given by \cref{eq:crossing_reduced} with $L=2$ as usual.

    In~\cite{1804.00949}, solutions to the SCWI were found in the form of polynomials of degree up to 10 in $s,t$. To write these, it is convenient to note that any crossing-symmetric function $f(s,t,\sigma,\tau)$ can be written in the form:
    \begin{align}
        \begin{split}
            f(s,t;\sigma,\tau) ={}& (1+\sigma^2+\tau^2)f_1 + (s+u\sigma^2+t\tau^2)f_2 + (s^2+u^2\sigma^2+t^2\tau^2)f_3\\
            &+ (\sigma+\tau+\sigma\tau)f_4 + (t\sigma + u\tau+s\sigma\tau)f_5 + (t^2\sigma+u^2\tau+s^2\sigma\tau)f_6\;.\label{eq:crossing_poly_ansatz}
        \end{split}
    \end{align}
    Where the $f_i$ are symmetric functions of $s,t,u$. In the case of polynomials, a convenient basis of such functions is given by combinations of $\sigma_2,\sigma_3$, defined as:
    \begin{equation}
        \sigma_2 = s^2+t^2+u^2,\quad \sigma_3=stu\;.
    \end{equation}
    The lowest-order polynomial is of degree 4, and is unique. Following~\cite{1804.00949}, we denote these polynomials as $M^{(n,m)}$ where $n$ is their order and $m$ counts separate polynomials. Therefore, the lowest one is $M^{(4,1)}$, and it is given by:
    \begin{align}
        \begin{split}
            f_1^{(4, 1)} &=\frac{\sigma_2^2}{4}  + \frac{6}{7} \sigma_3 - \frac{22}{5}\sigma_2 + \frac{96}{5}\;,\\
            f_2^{(4, 1)} &=  \sigma_3 + 2 
            \sigma_2 - \frac{736}{35}\;,\\
            f_3^{(4, 1)} &=  -\frac{\sigma_2}{2} + \frac{228}{35}\;,\\
            f_4^{(4, 1)} &=- \frac{104}{7} \sigma_3 - \frac{40}{7} \sigma_2 + \frac{4672}{35}\;,\\
            f_5^{(4, 1)} &=  2 \sigma_3 - \frac{18}{7} \sigma_2 - \frac{496}{7}\;,\\
            f_6^{(4, 1)} &=  \frac{832}{35}\;.
        \end{split}\label{eq:3d_deg_4}
    \end{align}
    Its primitive is then simply:
    \begin{equation}
        \widetilde{M}^{(4,1)} = \frac{1}{35}\;.
    \end{equation}
    The primitives of higher-degree polynomials can be found in \cref{sec:3d_higher_prim}. We note in particular how simple the primitives are compared to the corresponding polynomial amplitudes, due to the fact that there is only one function in the primitive, which is of 4 orders lower than the polynomial amplitudes it reproduces.

    \subsubsection{Supergravity}
    We now turn our attention to the supergravity amplitude. Up to a normalization factor, it is given by~\cite{1712.02800,2006.12505}:
    \begin{align}
        \begin{split}
            &\mathcal{M}_{\text{sugra}}(s,t;\sigma,\tau) = \mathcal{M}_{\text{sugra},s}(s,t,\sigma,\tau)+\mathcal{M}_{\text{sugra},t}(s,t,\sigma,\tau)+\mathcal{M}_{\text{sugra},u}(s,t,\sigma,\tau)\;,\\
            &\mathcal{M}_{\text{sugra},s}(s,t;\sigma,\tau) = \sum_{m=0}^\infty -\frac{(t-2)(u-2)+(s+2)((t-2)\sigma+(u-2)\tau)}{\Gamma(\frac{1}{2}-m)^2m!\Gamma(m+\frac{5}{2})(s-1-2m)}\;,\\
            &\mathcal{M}_{\text{sugra},t}(s,t,\sigma,\tau) = \tau^2\mathcal{M}_{\text{sugra},s}\left(t,s,\tfrac{\sigma}{\tau},\tfrac{1}{\tau}\right),\quad \mathcal{M}_{\text{sugra},u}(s,t,\sigma,\tau)=\mathcal{M}_{\text{sugra},s}\left(u,t,\tfrac{1}{\sigma},\tfrac{\tau}{\sigma}\right)\;.\label{eq:3d_sugra}
        \end{split}
    \end{align}
    The sum in $\mathcal{M}_{\text{sugra},s}$ can be performed, giving:
    \begin{equation}
        \mathcal{M}_{\text{sugra},s}(s,t;\sigma,\tau) = -\Big((t-2)(u-2)+(s+2)((t-2)\sigma+(u-2)\tau)\Big)\left(\frac{s+4}{2 \sqrt{\pi}s (s+2)}+\frac{4\Gamma\left(\frac{1}{2}-\frac{s}{2}\right)}{\pi s^2(s+2) \Gamma\left(-\frac{s}{2}\right)}\right)\;.
    \end{equation}
    We separate this into the rational and $\Gamma$-function parts, $\mathcal{M}_{\text{sugra},s}=\mathcal{M}_{\text{rat},s} +\mathcal{M}_{\Gamma,s}$, with:
    \begin{equation}
        \mathcal{M}_{\Gamma,s}(s,t;\sigma,\tau) = -\Big((t-2)(u-2)+(s+2)((t-2)\sigma+(u-2)\tau)\Big)\left(\frac{4\Gamma\left(\frac{1}{2}-\frac{s}{2}\right)}{\pi s^2(s+2) \Gamma\left(-\frac{s}{2}\right)}\right)\;,\label{eq:gamma_s}
    \end{equation}
    and $\mathcal{M}_{\text{rat},s}$ containing the rest, which is a rational function. Then, we can combine the rational parts of the three channels:
    \begin{equation}
        \mathcal{M}_{\text{rat}}(s,t;\sigma,\tau) = \mathcal{M}_{\text{rat},s}(s,t;\sigma,\tau)+\mathcal{M}_{\text{rat},t}(s,t;\sigma,\tau)+\mathcal{M}_{\text{rat},u}(s,t;\sigma,\tau)\;,
    \end{equation}
    where $\mathcal{M}_{\text{rat},t}$ and $\mathcal{M}_{\text{rat},u}$ are defined as in \cref{eq:3d_sugra}. We can also define $\mathcal{M}_{\Gamma,t}$ and $\mathcal{M}_{\Gamma,u}$ similarly. The advantage of this rewriting is that we can now write the full supergravity amplitude as a sum of four terms:
    \begin{equation}
        \mathcal{M}_{\text{sugra}}(s,t;\sigma,\tau) = \mathcal{M}_{\text{rat}}(s,t;\sigma,\tau) + \mathcal{M}_{\Gamma,s}(s,t;\sigma,\tau)+\mathcal{M}_{\Gamma,t}(s,t;\sigma,\tau)+\mathcal{M}_{\Gamma,u}(s,t;\sigma,\tau)
    \end{equation}
    where each of these terms solves the SCWI separately, such that we can individually find primitives for them. Furthermore, three of these terms are related by crossing symmetry, such that we need only find one primitive, and use crossing transformations in the reduced amplitude to find the others. By far, the simplest primitive is the rational one:
    \begin{equation}
        \widetilde{\mathcal{M}}_{\text{rat}}(s,t;\sigma,\tau) = \frac{3\tilde\sigma_3-2\tilde\sigma_2+16}{6\sqrt{\pi}s(s+2)t(t+2)\tilde u (\tilde u+2)}\;,
    \end{equation}
    where we introduced the crossing-symmetric basis polynomials:
    \begin{equation}
        \tilde\sigma_3 = st\tilde u,\quad \tilde\sigma_2 = s^2+t^2+\tilde u^2\;.\label{eq:sigma_pol_reduced}
    \end{equation}
    Finding the primitives for the other parts is more complicated. In fact, we have only been able to find them in certain limits, but we believe that this is strong evidence of their existence. The limits in which we gives expansions of the primitive are large $s$ (keeping $t$ fixed), large $t$ (keeping $s$ fixed) and large $s$ and $t$ (keeping $z\equiv \frac{s}{t}$ fixed).

    For all of these limits, we are solving the equation:
    \begin{equation}
        \mathcal{M}_{\Gamma,s}(s,t;\sigma,\tau) - \mathcal{D}\widetilde{\mathcal{M}}(s,t)=0\;.\label{eq:6d_sugra_diff}
    \end{equation}
    Taking the coefficients of $\sigma,\tau$ separately, this breaks down to six difference equations, which can be considered separately. We find that the simplest parametrization to solve these equations is to define $s=2S$, $t=2T$, and:
    \begin{equation}
        \widetilde{\mathcal{M}}(s,t) = \frac{\Gamma(\frac{1}{2}-S)}{\Gamma(1-S)}F(S,T)\;.
    \end{equation}
    This allows us to divide out by the $\Gamma$ function dependence, and thus solve rational difference equations of two variables. Then, taking $S$ or $T$ large reduces these to equations of one variable, while the large $S,T$ limit turns this into a differential equation problem. We find that in the large-$S$ limit, keeping $T$ fixed, $F(S,T)$ is:
    \begin{equation}
        F(S,T) = -\frac{1}{18\pi S^2} + \frac{8T+17}{90\pi S^3} - \frac{72T^2+288T+305}{630\pi S^4}+\ldots\;.
    \end{equation}
    In the opposite limit of large $T$ with $S$ kept fixed, we find instead:
    \begin{align}
        \begin{split}
        F(S,T) ={}& \frac{-1+24\pi A +H(S+1)-\log T}{24\pi T^2}\\
        &\frac{48 \pi  (A+B) S+48 \pi B-2 (S+2) H(S+2)+2 (S+2) \log(T)+3 S+6}{48 \pi  T^3} + \ldots\;,
        \end{split}
    \end{align}
    where $A$ and $B$ are constants and $H(x)$ is the harmonic number function. Higher-order terms do not contain more constants than $A$ and $B$.

    Finally, we consider the limit of large $S$ and $T$, keeping $z\equiv \frac{S}{T}$ fixed. In this limit, $F$ is a function of $T$ and $z$, and at each order in $T$, \cref{eq:6d_sugra_diff} becomes 6 ODEs in $z$. We decompose $F(T,z)$ as:
    \begin{equation}
        F(T,z) = T^{-2}\left(F_0(z) + \frac{F_1(z)}{T} + \frac{F_2(z)}{T^2}+\ldots\right)\;.
    \end{equation}
    Then, we can combine the ODEs to eliminate all of the $F_i(z)$ except $F_0$, expressing all of the others are derivatives of this one. For example, we get:
    \begin{equation}
        F_1(z) = \frac{8 \pi  z (z+1) \left((z+4)F_0'(z)+z (z+1) F_0''(z)\right)-2 \pi z (7 z+16) F_0(z)+1}{32 \pi  z (z+1)}\;.
    \end{equation}
    Then, we obtain a single ODE for $F_0(z)$, which can be solved:
    \begin{equation}
        F_0(z) = \frac{24 \pi  \widehat{A}+z \left(24 \pi  (\widehat{B}+\widehat{C}z)+\sqrt{z+1}\right)-z (z+2) \tanh^{-1}\left(\sqrt{z+1}\right)}{24 \pi z (z+1)^{3/2}}\;,
    \end{equation}
    where $\widehat{A}$, $\widehat{B}$ and $\widehat{C}$ are constants of integration. Finding these, as well as $A,B$ from the large $T$ expansion, requires solving a shooting problem. It may be possible to fix some (or all) of these constants by demanding matching of this last expansion to the large $S$ and large $T$ expansions in the appropriate limit of $z$, but it is not immediately clear that the order of these limits should be interchangeable, so we will not do this here.

    Finally, we note that $\widetilde{\mathcal{M}}_{\Gamma,t}(s,t)$ and $\widetilde{\mathcal{M}}_{\Gamma,u}(s,t)$ can be obtained from crossing symmetry:
    \begin{equation}
        \widetilde{\mathcal{M}}_{\Gamma,t}(s,t) = \widetilde{\mathcal{M}}_{\Gamma,s}(t,s),\quad \widetilde{\mathcal{M}}_{\Gamma,u}(s,t) = \widetilde{\mathcal{M}}_{\Gamma,s}(\tilde u,t)\;.
    \end{equation}
    
    \subsection{4d \texorpdfstring{$\mathcal{N}=4$}{N=4} SYM with line defects}\label{sec:defects}
    In this section, we consider 4d $\mathcal{N}=4$ SYM with the insertion of defects. Recently, Alday and Zhou found solutions to the Ward identities in this setting, both a supergravity amplitude and some polynomials~\cite{2411.04378}. We will show that these can be obtained from reduced Mellin amplitudes, as in the defect-free case. To begin, we will summarize the setting they used, and show how to obtain the SCWI in this context, following their approach.

    \subsubsection{Obtaining the SCWI with defects}

    We are interested in a line defect, dual to a D1 brane which occupies an $AdS_2$ subset of $AdS_5\times S^5$. We will be interested in the two-point function of operators in the stress tensor multiplet, in the presence of the half-BPS line defect. We can define such operators by:
    \begin{equation}
        S(x,u) = u^mu^n N_S\left(\phi^I_m(x)\phi^I_n(x)-\frac{1}{6}\delta_{mn}\phi^I_2(x)\phi^I_2(x)\right)\;,
    \end{equation}
    where $u^m$ is a null polarization vector of the SO(6) R-symmetry group and $I=1,\ldots,N^2-1$ is an adjoint index of the gauge group $SU(N)$. The normalization is:
    \begin{equation}
        N_S = \frac{2\sqrt{2}\pi^2}{g^2_{YM}\sqrt{N^2-1}}\;,
    \end{equation}
    such that the defect-free two-point function is:
    \begin{equation}
        \expval{S(x_1,u_1)S(x_2,u_2)} = \frac{(u_1\cdot u_2)^2}{x_{12}^4}\;.
    \end{equation}
    We now consider the same two-point function in the presence of a defect operator, which we denote by double angular brackets. We define:
    \begin{equation}
        \llangle S(x_1,u_1)S(x_2,u_2)\rrangle = \frac{(u_1\cdot\theta)^2(u_2\cdot\theta)^2}{|x_1^i|^2|x_2^i|^2}\mathcal{F}(\zeta,\eta;\sigma)\;,
    \end{equation}
    where $\theta$ is a fixed unit vector, which captures the R-symmetry breaking from $SO(6)$ to $SO(5)$ by the line defect. If we call $\tau$ the coordinate parallel to the defect and denote by $x^{i=2,\ldots,4}$ the transverse coordinates, the relevant position cross-ratios are:
    \begin{equation}
        \zeta = \frac{(\tau_1-\tau_2)^2+|x_1^i|^2+|x_2^i|^2}{2|x_1^i||x_2^i|}\equiv \frac{\xi+\chi}{2},\quad \eta = \frac{x_1^jx_2^j}{|x_1^i||x_2^i|}\equiv \frac{\chi}{2}\;,
    \end{equation}
    with the $R$-symmetry cross-ratio given by:
    \begin{equation}
        \sigma = \frac{(u_1\cdot u_2)}{(u_1\cdot\theta)(u_2\cdot\theta)}\;.
    \end{equation}
    As in the defect-free case, the relevant SCWI are given by the action of the fermionic generators, and take the form~\cite{1608.05126}:
    \begin{equation}
        \eval{\left(\partial_z+\frac{1}{2}\partial_\alpha\right)\mathcal{F}(z,\bar z,\alpha)}_{\alpha=z}=0\;,
    \end{equation}
    and another one with $z\leftrightarrow\bar z$, where $z,\bar{z}$ are defined as:
    \begin{equation}
        \zeta = \frac{1+z\bar z}{2\sqrt{z\bar z}},\quad \eta = \frac{z+\bar z}{2\sqrt{z\bar z}},\quad \sigma = -\frac{(1-\alpha)^2}{2\alpha}\;.
    \end{equation}
    We can interpret these coordinates physically by using conformal symmetry to move the operators to a 2d plane:
    \begin{equation}
        x_1 = \left(\frac{z-\bar z}{2i},\frac{z+\bar z}{2},0,0\right),\quad x_2 = (0,1,0,0)\;.
    \end{equation}
    We place the line defect at $x^2=x^3=x^4=0$. Then, $\mathcal{F}(z,z;z)$ only depends on the sign of $z$, such that it takes on two values, depending on whether the two operators are on the same side of the defect in this plane, or opposite ones~\cite{2411.04378}.

    To translate the Ward identities to Mellin space, we follow~\cite{2306.11896,2411.04378}. We define:
    \begin{equation}
        \mathbb{W}(z,\bar z) = \eval{\left(\partial_z+\frac{1}{2}\partial_\alpha\right)\mathcal{F}(z,\bar z,\alpha)}_{\alpha=z}\;,
    \end{equation}
    which must vanish by the SCWI. We then separate this into four equations, in two steps. First, define:
    \begin{equation}
        \mathbb{W}_+(z,\bar z) = \mathbb{W}(z,\bar z)+\mathbb{W}(1/z,1/\bar z),\quad \mathbb{W}_-(z,\bar z) = \frac{1-z^2}{2z}(\mathbb{W}(z,\bar z)-\mathbb{W}(1/z,1/\bar z))\;,
    \end{equation}
    which we then use to create four independent linear combinations:
    \begin{align}
        \begin{split}
            \mathbb{W}_{++}(z,\bar z) = \mathbb{W}_+(z,\bar z)+\mathbb{W}_+(\bar z,z),\quad \mathbb{W}_{+-}(z,\bar z) = \frac{z\bar z}{(\bar z-z)(1-z\bar z)}(\mathbb{W}_+(z,\bar z) -\mathbb{W}_+(\bar z,z))\;,\\
            \mathbb{W}_{-+}(z,\bar z) = \mathbb{W}_-(z,\bar z)+\mathbb{W}_-(\bar z,z),\quad \mathbb{W}_{--}(z,\bar z) = \frac{z\bar z}{(\bar z-z)(1-z\bar z)}(\mathbb{W}_-(z,\bar z) -\mathbb{W}_-(\bar z,z))\;.
        \end{split}
    \end{align}
    Each of these will give a different SCWI equation in Mellin space, for four total equations. To translate them, we start by rewriting the derivatives in terms of the cross-ratios $\xi,\chi$. For a general 2-point function $\llangle S_1S_2\rrangle$ where the operators respectively have conformal dimensions $\Delta_1,\Delta_2$, we can expand the two-point as a polynomial in $\sigma$ of degree $\Delta_m=\min\{\Delta_1,\Delta_2\}$:
    \begin{equation}
        \mathcal{F}(\xi,\chi;\sigma) = \sum_{j=0}^{\Delta_m}\sigma^j\mathcal{F}_j(\xi,\chi)\;.
    \end{equation}
    In this language, $\mathbb{W}$ is given by:
    \begin{equation}
        \mathbb{W}(z,\bar{z})=\sum_{j=0}^{\Delta_m}\frac{(1-z)^{2j}}{z^j}\left(\left(\frac{\bar{z}-1}{\sqrt{z\bar{z}}}-\frac{\xi}{2z}\right)\partial_\xi+\left(\frac{1}{\sqrt{z\bar{z}}}-\frac{\chi}{2z}\right)\partial_\chi+\frac{j(z+1)}{2z(z-1)}\right)\frac{\mathcal{F}_j(\xi,\chi)}{(-2)^j}\;.
    \end{equation}
    Then, $\mathbb{W}_\pm$ have the representations:
    \begin{align}
        \begin{split}
            \mathbb{W}_{+}(z,\bar{z})={}&\sum_{j=0}^{\Delta_m}\frac{(1-z)^{2j}}{z^j}\bigg(-\left(\frac{(1-z)^2}{2z}+2\right)\xi\partial_\xi+\left(\xi-\frac{(1-z)^2}{2z}\chi\right)\partial_\chi\\
            {}&\quad\quad-\frac{j(1-z)^2}{2z}-2j\bigg)\frac{\mathcal{F}_j(\xi,\chi)}{(-2)^j}\;,\\
            \mathbb{W}_{-}(z,\bar{z})={}&\sum_{j=0}^{\Delta_m}\frac{(1-z)^{2j}}{z^j}\bigg(-\left(\left(\frac{(1-z)^2}{2z}\right)^2+\frac{3(1-z)^2}{2z}+2\right)\xi\partial_\xi\\
            {}&\quad\quad+\left(\left(\frac{(1-z)^2}{2z}+1\right)\xi-\left(\left(\frac{(1-z)^2}{2z}\right)^2+\frac{(1-z)^2}{2z}\right)\chi\right)\partial_\chi\\
            {}&\quad\quad-j\left(\frac{(1-z)^2}{2z}\right)^2-\frac{3j(1-z)^2}{2z}-2j\bigg)\frac{\mathcal{F}_j(\xi,\chi)}{(-2)^j}\;.
        \end{split}
    \end{align}
    From these, we form the $\mathbb{W}_{\pm\pm}$, which can be written in terms of the cross-ratio combinations~\cite{2411.04378}:
    \begin{equation}
        \zeta^{(j)}_+=\frac{(1-z)^{2j}}{2z^j}+\frac{(1-\bar{z})^{2j}}{2\bar{z}^j}\;,\quad \zeta^{(j)}_-=\frac{z\bar{z}}{(\bar{z}-z)(1-z\bar{z})}\left(\frac{(1-z)^{2j}}{2z^j}-\frac{(1-\bar{z})^{2j}}{2\bar{z}^j}\right)\;.
    \end{equation}
    These can be rewritten as polynomials in the $\xi,\chi$ variables~\cite{2306.11896}:
    \begin{align}
        \begin{split}
            \zeta_{+}^{(i)}
            & = \sum _{j=0}^{\left\lfloor i/2\right\rfloor } \binom{i}{2 j}
            \left(\frac{\chi^2}{4}-1\right)^j \left((\chi +\xi )^2-4\right)^j
            \left(\frac{\chi^2}{2} +\frac{\xi\chi}{2}-2\right)^{i-2 j}\;,\\
            \zeta_{-}^{(i)}
            & = \sum _{j=0}^{\left\lfloor i/2\right\rfloor } \binom{i}{2 j+1}
            \left(\frac{\chi^2}{4}-1\right)^j \left((\chi+\xi )^2-4\right)^j
            \left(\frac{\chi^2}{2} +\frac{\xi\chi}{2}-2\right)^{i-2 j-1}\;.
        \end{split}
    \end{align}
    Now, we are ready to translate the SCWI into Mellin space. Mellin amplitudes for defect CFTs were introduced in~\cite{1705.05362,1803.06721}. We'll use the conventions of~\cite{2411.04378}:
    \begin{equation}
        \mathcal{F}(\xi,\chi) = C_{\Delta_1\Delta_2}\int\frac{\dd\delta\dd\gamma}{(2\pi i)^2}\xi^{-\delta}\chi^{-\gamma+\delta}\mathcal{M}(\delta,\gamma)\Gamma(\delta)\Gamma(\gamma-\delta)\prod_{i=1}^2\Gamma\left(\frac{\Delta_i-\gamma}{2}\right)\;.
    \end{equation}
    with the normalization:
    \begin{equation}
        C_{\Delta_1\Delta_2} = \frac{\pi^{1/2}\Gamma\left(\frac{\Delta_1+\Delta_2-1}{2}\right)}{4\Gamma(\Delta_1)\Gamma(\Delta_2)}\;.
    \end{equation}
    With these conventions, we can translate the SCWI to Mellin space with the following dictionary:
    \begin{align}
        \begin{split}
            \xi\partial_\xi\to (-\delta)\mathds{1}\;,\quad& \chi\partial_\chi\to (-\gamma+\delta)\mathds{1}\;,\\
            \xi^m\chi^n\to (\delta)_m(\gamma-\delta)_n&\prod_{i=1}^2\left(\frac{\Delta_i-\gamma}{2}\right)_{-\frac{m+n}{2}}P_{m,m+n}\;,
        \end{split}
    \end{align}
    where now the shift operators act on functions of $\delta,\gamma$:
    \begin{equation}
        P_{m,n}f(\delta,\gamma) = f(\delta+m,\gamma+n)\;.
    \end{equation}
    Using this dictionary, we get four independent SCWI equations from setting the $\mathbb{W}_{\pm\pm}$ to 0. Once again, these are difference relations, as we had in the previous sections.

    \subsubsection{Solutions to the SCWI}
    The SCWI of the previous section can be solved by using the techniques of \cref{sec:mellin_sol}. We now specialize to the stress tensor multiplet. In this case, our method naturally finds two operators, with separate ambiguity functions. However, for a specific choice of these functions, we get an operator $\mathcal{D}$ with minimal shifts, and we can recover the original functions by right-acting on $\mathcal{D}$ with the appropriate operators, such that there is only one fundamental solution to consider. This time, the SCWI involve shifts in the positive $\delta,\gamma$ direction, so we choose to define our operator in the same way, which is, as before, simply a choice of a shift in the ambiguity. We separate the operator into three components as:
    \begin{equation}
        \mathcal{D} = \sum_{i=0}^2\sigma^i\mathcal{D}_i\;.
    \end{equation}
    Then, the individual components are given by the following:
    \begin{align}
        \begin{split}
        \mathcal{D}_0 ={}& -(\gamma -2) (\delta +1) P_{2,2}+(\delta +1) (\gamma
        -\delta +1)P_{2,4}+\frac{1}{2} (\gamma -2) \delta 
        P_{3,2}+\frac{1}{2}
        \delta  (-\gamma +2 \delta +1)
        P_{3,4}\;,\\
        \mathcal{D}_1 ={}&\frac{(\gamma -2) \left(2 \gamma ^2-6
        \gamma  \delta +3 \gamma +3 \delta
        ^2+\delta -4\right)}{\delta
        -1}P_{1,2}-\frac{(\gamma -\delta )
        \left(\gamma ^2-5 \gamma  \delta +7
        \gamma +4 \delta ^2-12 \delta
        +8\right)}{\delta -1}P_{1,4}\\
        &+\frac{2 (\gamma -4)(\gamma -2)^2}{\delta -1}P_{0,0}-\frac{(\gamma -4)(\gamma -2)^2 }{\delta -1}P_{1,0}-\frac{2 (\gamma -2)(\gamma -\delta +1) (2 \gamma -\delta-2)}{\delta-1}P_{0,2}\\
        &+(\gamma-2) (\gamma -\delta +1)
        P_{2,2}+\frac{2(\gamma -\delta ) (\gamma -\delta +1)(\gamma -\delta +3)}{\delta -1}P_{0,4}-(\gamma -2 \delta +2)(\gamma -\delta )P_{2,4}\;,\\
        \mathcal{D}_2 ={}& -\frac{(\gamma -4) (\gamma -2)^2}{\delta-1}P_{0,0}+\frac{(\gamma -4) (\gamma -2)^2}{2 (\delta-1)}P_{1,0}+\frac{(\gamma -3) (\gamma -2)(\gamma -\delta +1)}{\delta -1}P_{0,2}\\
        &-\frac{(\gamma -3)(\gamma -2) (\gamma -2 \delta +2)}{2 (\delta-1)}P_{1,2}\;.
        \end{split}
    \end{align}
    While we lack the crossing symmetry we had in the other theories, the ambiguities were fixed here by demanding that the polynomial solutions to the SCWI of~\cite{2411.04378} should have polynomial primitives. For example, the first of these polynomial solutions is of order 2 in $\delta,\gamma$, and given by:
    \begin{equation}
        M_{2}^{(1)}=\frac{1}{5} \left(5 \delta ^2+11 \delta +12-2 \sigma  (5 \gamma  \delta -32 \gamma +8 \delta +28)+(\gamma -2) (5 \gamma -14) \sigma ^2\right)\;,
    \end{equation}
    up to a normalization factor. Its primitive is then:
    \begin{equation}
        \widetilde{M}_{L=2}^{(1)} = \frac{2}{5}\delta\;.\label{eq:defect_ord_2_prim}
    \end{equation}
    Primitives for the higher-order polynomials are given in \cref{sec:defect_higher_prim}. It should be noted that, due to the denominators in the operator, it is not true that any polynomial primitive has a polynomial image. For example, we find:
    \begin{equation}
        \mathcal{D}(1) = \frac{3}{2}(\delta+2) + \sigma\left(\frac{4 (3 \gamma -4)}{\delta -1}-3 \gamma -2\right)+ \sigma^2 \frac{(\gamma -2) (3 \gamma -8)}{2 (\delta -1)}\;.
    \end{equation}
    However, in trying all polynomials up to degree 6 in $\delta,\gamma$ which solve the SCWI, it is the case that this operator always has a polynomial primitive for them. If we were to demand instead that any polynomial be taken to a polynomial, and fix the ambiguity with this requirement, polynomial solutions to the SCWI would in general have rational primitives.

    We can give a method for building functions which will transform into polynomials under the action of $\mathcal{D}$. In order to cancel the poles at $\delta=1$ in the operator, it suffices to build functions out of powers of $\gamma$ and the shifted variables:
    \begin{equation}
        d_n \equiv (\delta^n+2^n-2)\;.\label{eq:defect_dn}
    \end{equation}
    Linear combinations of terms of the form $\gamma^md_n$ with $n>0$ will transform into polynomials, and terms left over without any $d_n$ after rewriting a polynomial in $\delta,\gamma$ in terms of these variables will contribute to a rational function. Since $\delta=d_1$, transforming the first primitive above into this language is trivial.

    We also want to obtain the primitive for the supergravity amplitude, which was also given in~\cite{2411.04378}. Up to a normalization factor, it is:
    \begin{align}
        \begin{split}
            \mathcal{M}_{\text{sugra}}={}&\frac{\sigma  (\sigma  (\gamma -\delta )-\sigma +4)}{\delta -1}-\frac{2 (\gamma -\delta )+4 \sigma -4}{\gamma }+\frac{\gamma -\delta }{\gamma +2}+(\sigma -1)^2\\
            {}&+\frac{2^{\gamma +2} \Gamma (-\gamma ) (\frac{\delta +2}{\gamma +2}-\sigma)}{\Gamma (\frac{2-\gamma }{2})^2}\;,
        \end{split}
    \end{align}
    for which we find the primitive:
    \begin{equation}
        \widetilde{\mathcal{M}}_{\text{sugra}} = \frac{2}{\gamma -2}+\frac{(\gamma-3)2^{\gamma}\Gamma(2-\gamma)}{(\gamma-4)\Gamma\left(2-\frac{\gamma }{2}\right)^2}\;.
    \end{equation}

    \section{Conclusion}
    \label{sec:conclusion}
    In this paper, we have developed a method to solve the SCWI directly in Mellin space, without referring to a solution in position space. This approach is based on using functions which diagonalize the shift operators to extract an operator $\mathcal{D}$ such that:
    \begin{equation}
        \mathcal{M}(s,t;\sigma,\tau) = \mathcal{D}\widetilde{\mathcal{M}}(s,t)
    \end{equation}
    in the simplest case, considered in most of this paper, which is that of identical particles with $k_i=2$. $\mathcal{D}$ is built for this to hold for any arbitrary function $\widetilde{\mathcal{M}}(s,t)$. In more complicated cases, such as the $k_i=3$ case of \cref{sec:ki_3}, the reduced Mellin amplitude is also allowed to depend on $\sigma,\tau$, but in a simpler way than the full amplitude. We also consider a simple mixed correlator in \cref{sec:22kk} in 4d $\mathcal{N}=4$ SYM.

    Mathematically, if we consider the SCWI to be given by operators $\mathcal{W}_i$, such that the SCWI equations are, as in \cref{eq:scwi_mellin_op}, given by $\mathcal{W}_if(s,t)=0$, we would say that $\mathcal{D}$ is a projector into the kernel of the $\mathcal{W}_i$ operators. We constructed this projector explicitly in the cases of 4d $\mathcal{N}=4$ SYM, 6d $\mathcal{N}=(2,0)$ theory, 3d $\mathcal{N}=8$ ABJM and 4d $\mathcal{N}=4$ SYM with line defects, with the last two being new results. We also explained how to derive the modified constraints of crossing symmetry, and gave them in each theory where it was applicable.

    We have not proven in this work that all solutions to the SCWI take the form $\mathcal{D}\widetilde{\mathcal{M}}(s,t)$, but we have shown that an extensive number of results from the literature can be rewritten in terms of an appropriate reduced Mellin amplitude acted upon by $\mathcal{D}$, in each of the new theories, which we believe to be evidence that many physically relevant amplitudes can be written in this language. We would hope that a proof of the full claim might be done in future work.

    There are of course still many correlators for which we have not explicitly solved the SCWI, such as equal correlators with $k_i\ge 4$ or other types of mixed correlators than those of \cref{sec:22kk}. Since the operator $\mathcal{D}$ takes the same form in position space in 4d $\mathcal{N}=4$ SYM~\cite{1608.06624,1710.05923}, and for equal-operator correlators in 6d $\mathcal{N}=(2,0)$ theory~\cite{1712.02788}, we expect that our method of solving the SCWI would lead to fruitful results in those cases as well.

    The most natural application of this work is to bootstrap the theories of 3d $\mathcal{N}=8$ ABJM and 4d $\mathcal{N}=4$ SYM with line defect insertions, using reduced amplitudes, in order to understand new correlators in these theories. It would also be interesting to see if the SCWI for higher-point functions can be solved by an extension of our techniques here. It may finally also be possible that our techniques here could be adapted to solve the SCWI in other theories, either further defect theories or theories with less supersymmetry.

    \section*{Acknowledgements}
    We would like to thank L.F. Alday for helpful discussions, as well as X. Zhou, M. Nocchi and V. Pellizzani for insightful comments on the manuscript. The work of CV is supported by the Fonds de recherche du Qu\'ebec, secteur Nature et technologies, through its Doctoral Training
    Scholarships program, scholarship number 349150. 
    
    For the purpose of open access, the author has applied a CC BY public copyright licence to any Author Accepted Manuscript (AAM) version arising from this submission.

    \appendix
    \section{Full operators}\label{sec:full_ops}
    \subsection{6d \texorpdfstring{$\mathcal{N}=(2,0)$}{N=(2,0)} theory full operator}\label{sec:6d_full_op}
    In this appendix, we present the full operator for the 6d $\mathcal{N}=(2,0)$ theory SCWI. The $\mathcal{D}_{0,0}$ component was already given in \cref{eq:6d_d_00}, and we complete the operator below:
    \begin{align}
        \begin{split}
            \mathcal{D}_{0,1} ={}& (s-10) (s-8)^2 (s+t-10) (s+t-8)P_{-4,0}\\
            &+(s-8)(t-8)(s+t-12)(s+t-10)(s+t-8)P_{-2,-2}\\
            &-(s-8)(s+t-8)^2 (s+t-6)(2 s+t-16)P_{-2,0}\\
            &+(t-10)(t-8)^2 (s+t-10) (s+t-8)P_{0,-4}\\
            &-(t-8) (s+2 (t-8))(s+t-8)^2 (s+t-6)P_{0,-2}\\
            &+(s+t-8)^2 (s+t-6)^2(s+t-4)P_{0,0}\;,
        \end{split}
    \end{align}
    \begin{align}
        \begin{split}
            \mathcal{D}_{0,2} = {}&-(s-10) (s-8)^2 (s+t-10) (s+t-8)
            P_{-4,0}\\
            &-(s-6) (s-8) (t-8) (s+t-10) (s+t-8)
            P_{-2,-2}\\
            &+(s-6) (s-8) (s+t-8)^2 (s+t-6)P_{-2,0}\;,
        \end{split}
    \end{align}
    \begin{align}
        \begin{split}
            \mathcal{D}_{1,0}={}& (s-10) (s-8)^2 (t-8) (t-6)P_{-4,-2}\\
            &+(s-8)(t-10)(t-8)^2 (s-t)P_{-2,-4}\\
            &+(s-8)(t-8) (t-6) (t-4)(s+t-8)P_{-2,-2}\\
            &-(t-12)(t-10)^2 (t-8)^2P_{0,-6}\\
            &+(t-10) (t-8)^2 (s+2(t-8))(s+t-8)P_{0,-4}\\
            &-(t-8) (t-6) (s+t-8)^2 (s+t-6)P_{0,-2}\;,
        \end{split}
    \end{align}
    \begin{align}
        \begin{split}
            \mathcal{D}_{1,1}={}&-(s-12) (s-10)^2 (s-8)^2P_{-6,0}\\
            &-(s-10)(s-8)^2 (t-8) (s-t)P_{-4,-2}\\
            &+(s-10) (s-8)^2(s+t-8) (2 s+t-16)P_{-4,0}\\
            &+(s-6) (s-8) (t-10)(t-8)^2P_{-2,-4}\\
            &+(s-6) (s-4) (s-8) (t-8) (s+t-8)P_{-2,-2}\\
            &-(s-6) (s-8) (s+t-8)^2 (s+t-6)P_{-2,0}\;,
        \end{split}
    \end{align}
    \begin{align}
        \begin{split}
            \mathcal{D}_{2,0}={}&-(s-10) (s-8)^2 (t-8) (t-6)P_{-4,-2}\\
            &-(s-6) (s-8) (t-10) (t-8)^2P_{-2,-4}\\
            &+(s-6) (s-8) (t-8) (t-6) (s+t-8)P_{-2,-2}\;.
        \end{split}
    \end{align}

    \subsection{3d \texorpdfstring{$\mathcal{N}=8$}{N=8} ABJM full operator}\label{sec:3d_full_op}
    In this appendix, we present the full operator for the 3d $\mathcal{N}=8$ ABJM SCWI. The $\mathcal{D}_{0,0}$ component was already given in \cref{eq:3d_d_00}, and we complete the operator below:
    \begin{align}
        \begin{split}
            \mathcal{D}_{0,1} ={}& (s-4) (s-2) (s+t-2) (s+t) \left(3 s^2+4 s (t-2)+t^2-5t+4\right)P_{-4,0}\\
            &+(t-4) (t-2) (s+t-2) (s+t)\left(s^2+4 (s-2) t-5 s+3 t^2+4\right)P_{0,-4}\\
            &+(s-2) (t-2) (s+t-2) (s+t) \left(4 s^2+s (8t-15)+t (4 t-15)+16\right) P_{-2,-2}\\
            &-(s-6)(s-4) (s-2)^2 (s+t-2) (s+t-1)P_{-6,0}\\
            &-(s-4)(s-2) (t-2) (s+t-2) (s+t-1) (2 s+t-8)P_{-4,-2}\\
            &-(s-2) (t-4) (t-2) (s+t-2) (s+t-1) (s+2t-8)P_{-2,-4}\\
            &-(s-2) (s+t-2)^2 (s+t) (s+t+2) (3s+2 t-1)P_{-2,0}\\
            &-(t-6) (t-4) (t-2)^2 (s+t-2)(s+t-1)P_{0,-6}\\
            &-(t-2) (s+t-2)^2 (s+t) (s+t+2) (2s+3 t-1)P_{0,-2}\\
            &+(s+t-2)^2 (s+t)^2 (s+t+2) (s+t+4)P_{0,0}\;,
        \end{split}
    \end{align}

    \begin{align}
        \begin{split}
            \mathcal{D}_{0,2} ={}& (s-6) (s-4) (s-2)^2 (s+t-2) (s+t-1)P_{-6,0}\\
            &+(s-4) (2s-5) (s-2) (t-2) (s+t-2) (s+t-1)P_{-4,-2}\\&
            -(s-4)(s-2) (s+t-2) (s+t) (2 s (s+t-4)-5 t+5)P_{-4,0}\\
            &+(s-3) (s-2) (t-4) (t-2) (s+t-2) (s+t-1)P_{-2,-4}\\
            &-(s-3) (s-2) (t-2) (s+t-2) (s+t) (2 s+2t-3)P_{-2,-2}\\
            &+(s-3) (s-2) (s+t-2)^2 (s+t) (s+t+2)P_{-2,0}\;,
        \end{split}
    \end{align}

    \begin{align}
        \begin{split}
            \mathcal{D}_{1,0} ={}& (t-4) (t-2) (s+t-2) (s+t) \left(2 (s-8) t-5 (s-4)+3t^2\right)P_{0,-4}\\
            &-(s-6) (s-4) (s-2)^2 (t-3)(t-2)P_{-6,-2}\\
            &-(s-4) (s-2) (t-4) (t-2) (s(2 t-5)-(t-3) t)P_{-4,-4}\\
            &+(s-4) (s-2) (t-3) (t-2)(s-t-4) (s+t-2)P_{-4,-2}\\
            &-(s-2) (t-6) (t-4)(t-2)^2 (s-2 t+7)P_{-2,-6}\\
            &-(s-2) (t-4) (t-2) (t(4 t-17)+20) (s+t-2)P_{-2,-4}\\
            &+(s-2) (t-3) (t-2)(s+t-2) (s+t) (s+2 t)P_{-2,-2}\\
            &+(t-8) (t-6)(t-4)^2 (t-2)^2P_{0,-8}\\
            &-(t-6) (t-4) (t-2)^2(s+t-2) (s+3 t-11)P_{0,-6}\\
            &-(t-3) (t-2) (s+t-2)^2(s+t) (s+t+2)P_{0,-2}\;,
        \end{split}
    \end{align}

    \begin{align}
        \begin{split}
            \mathcal{D}_{1,1} ={}& (s-4) (s-2)\left(3 s^2+2 s (t-8)-5 (t-4)\right)(s+t-2) (s+t)P_{-4,0}\\
            &+(s-8) (s-6) (s-4)^2 (s-2)^2P_{-8,0}\\
            &+(s-6) (s-4) (s-2)^2 (t-2) (2 s-t-7)P_{-6,-2}\\
            &-(s-6) (s-4) (s-2)^2 (s+t-2) (3 s+t-11)P_{-6,0}\\
            &+(s-4) (s-2) (t-4) (t-2) (s (s-2 t-3)+5 t)P_{-4,-4}\\
            &-(s-4) (s (4 s-17)+20) (s-2) (t-2)(s+t-2)P_{-4,-2}\\
            &-(s-3) (s-2) (t-6) (t-4) (t-2)^2P_{-2,-6}\\
            &-(s-3) (s-2) (t-4) (t-2) (s-t+4) (s+t-2)P_{-2,-4}\\
            &+(s-3) (s-2) (t-2) (s+t-2) (s+t) (2 s+t)P_{-2,-2}\\
            &-(s-3) (s-2) (s+t-2)^2 (s+t) (s+t+2)P_{-2,0}\;,
        \end{split}
    \end{align}

    \begin{align}
        \begin{split}
            \mathcal{D}_{2,0} ={}& (s-6) (s-4) (s-2)^2 (t-3) (t-2)P_{-6,-2}\\
            &+(s-4) (s-2)(t-4) (t-2) (2 s t-5 s-5 t+15)P_{-4,-4}\\
            &-(s-4) (2s-5) (s-2) (t-3) (t-2) (s+t-2)P_{-4,-2}\\
            &+(s-3)(s-2) (t-6) (t-4) (t-2)^2P_{-2,-6}\\
            &-(s-3) (s-2)(t-4) (t-2) (2 t-5) (s+t-2)P_{-2,-4}\\
            &+(s-3)(s-2)(t-3)(t-2) (s+t-2) (s+t)P_{-2,-2}\;.
        \end{split}
    \end{align}

    \section{Higher-order primitives}\label{sec:higher_order_prim}
    \subsection{3d \texorpdfstring{$\mathcal{N}=8$}{N=8} ABJM}\label{sec:3d_higher_prim}
    We now consider the higher-order polynomial solutions of the SCWI for 3d $\mathcal{N}=8$ ABJM, which are known to degree 10 in the literature~\cite{1804.00949}. We already found a primitive for the degree 4 polynomial in \cref{eq:3d_deg_4}. As before, the primitives are linear combinations of terms of the form $\tilde\sigma_2^n\tilde\sigma_3^m$ (with $\tilde\sigma_i$ as defined in \cref{eq:sigma_pol_reduced}), which have degree $2n+3m$ in $s,t$, and correspond to polynomials of degree $2n+3m+4$ after the action of $\mathcal{D}$. This means that there can't be a new solution of degree 5, since we'd need a combination of degree 1 in the primitive, and indeed there is none in~\cite{1804.00949}. This also explains why there is exactly 1 new polynomial at each order $6,7,8,9$, and 2 new ones at order 10, since these correspond to orders $2,3,4,5$ and 6 for the primitive, respectively. For the polynomials themselves, we refer the reader to~\cite{1804.00949}, and we will only present the  primitives here. The order 6 polynomial, $M^{(6,1)}$, has the primitive:
    \begin{equation}
        \widetilde{M}^{(6,1)} = \frac{1}{99}\tilde\sigma_2 - \frac{32}{77}\;.
    \end{equation}
    The order 7 polynomial, $M^{(7,1)}$, has the primitive:
    \begin{equation}
        \widetilde{M}^{(7,1)} = \frac{1}{143}\tilde\sigma_3 + \frac{5}{39}\tilde\sigma_2 -\frac{768}{143}\;.
    \end{equation}
    The order 8 polynomial, $M^{(8,1)}$, has the primitive:
    \begin{equation}
        \widetilde{M}^{(8,1)} = \frac{1}{195}\tilde\sigma_2^2-\frac{64928}{177957}\tilde\sigma_2+\frac{72256}{1482975}\tilde\sigma_3+\frac{336256}{54925}\;.
    \end{equation}
    The order 9 polynomial, $M^{(9,1)}$, has the primitive:
    \begin{equation}
        \widetilde{M}^{(9,1)} = \frac{\tilde\sigma_3\tilde\sigma_2}{255}-\frac{3308 \tilde\sigma _2^2}{2740485}+\frac{100542256 \tilde\sigma_2}{1389425895}-\frac{64037048 \tilde\sigma_3}{463141965}-\frac{55950464}{92628393}\;.
    \end{equation}
    Finally, we reach order 10. There are two polynomials of this order, the first being denoted by $M^{(10,1)}$, with primitive:
    \begin{align}
        \begin{split}
            \widetilde{M}^{(10,1)} ={}& \frac{\tilde\sigma_2^3}{323}-\frac{2937984089344 \tilde\sigma_2^2}{11832185872815}-\frac{3828377589664 \tilde\sigma_3\tilde\sigma_2}{11832185872815}+\frac{25947413285888\tilde\sigma_2}{3944061957605}\\
            &+\frac{142788698967104\tilde\sigma_3}{11832185872815}-\frac{302986684604416}{2366437174563}\;.
        \end{split}
    \end{align}
    The other order 10 polynomial, $M^{(10,2)}$, has the primitive:
    \begin{align}
        \begin{split}
            \widetilde{M}^{(10,2)} ={}& \frac{\tilde\sigma_3^2}{323}-\frac{691086568 \tilde\sigma_2^2}{1690312267545}-\frac{4382344693 \tilde\sigma_3\tilde\sigma_2}{1690312267545}+\frac{13242609728 \tilde\sigma_2}{1690312267545}\\
            &+\frac{3458205064 \tilde\sigma_3}{99430133385}-\frac{9893699584}{338062453509}\;.
        \end{split}
    \end{align}

    \subsection{4d \texorpdfstring{$\mathcal{N}=4$}{N=4} with defects}\label{sec:defect_higher_prim}
    In~\cite{2411.04378}, polynomial solutions of the SCWI of order 2, 3 and 4 in $\delta,\gamma$ were found. The primitive for the order 2 polynomial was already given in \cref{eq:defect_ord_2_prim}. Here, we will present the higher-order cases. As before, we refer the reader to~\cite{2411.04378} for the full polynomials. There are two polynomials of degree 3, $M^{(i)}_3$, with primitives:
    \begin{align}
        \begin{split}
            \widetilde{M}_3^{(1)} ={}&\frac{2}{21}\delta(3\gamma-8) = \frac{2}{21}d_1(3\gamma-8)\;,\\
            \widetilde{M}_3^{(2)} ={}&\frac{2}{7}(\delta-1)(\delta-2) = \frac{2}{7}(d_2-3d_1)\;,
        \end{split}
    \end{align}
    where the $d_n$ variables are shifts of $\delta^n$ defined in \cref{eq:defect_dn}. Now, we consider the order-4 polynomials $M^{(i)}_4$, which have the primitives:
    \begin{align}
        \begin{split}
            \widetilde{M}^{(1)}_4 ={}&\frac{2}{135} \left(\delta  \left(15\gamma ^2-84 \gamma -4 \delta+116\right)-8\right) = \frac{1}{135} (6 \gamma  (5 \gamma -28) d_1+232
            d_1-8 d_2)\;,\\
            \widetilde{M}^{(2)}_4 ={}&\frac{2}{9} (\gamma -4)(\delta -2)(\delta -1) = \frac{2}{9}(\gamma-4)(d_2-3d_1)\;,\\
            \widetilde{M}^{(3)}_4 ={}&\frac{1}{9} (\delta -2)(\delta -1)(2\delta -3) = \frac{1}{9} (13 d_1-9 d_2+2 d_3)\;.
        \end{split}
    \end{align}

    \section{Some higher-\texorpdfstring{$k$}{k} solutions}\label{sec:more_correlators}
    \subsection{\texorpdfstring{$k_i=3$}{ki=3}}\label{sec:ki_3}
    We now consider the case of correlators of equal operators where $k_i=3$. We'll work only in the two theories where the answer is known from position space for the sake of simplicity and to be able to compare to previous results. However, the generalization here should also work for the other theories.

    \subsubsection{4d \texorpdfstring{$\mathcal{N}=4$}{N=4} SYM}
    In 4d $\mathcal{N}=4$ SYM, the operator from position space does not depend on the $\sigma,\tau$ indices of the function on which it acts. This means that we can still write down a single operator which works for any general function $\mathcal{M}(s,t;\sigma,\tau)$ which is linear in $\sigma,\tau$. We can thus still separate the operator as in \cref{eq:D_separation}, and write:
    \begin{align}
        \begin{split}
            \mathcal{D}_{0,0} ={}& (t-6)^2 (s+t-6)^2P_{0,-2}\;,\\
            \mathcal{D}_{0,1} ={}& (s+t-6)^2 \left(-(s-6)^2P_{-2,0}-(t-6)^2P_{0,-2}+(s+t-4)^2P_{0,0}\right)\;,\\
            \mathcal{D}_{0,2} ={}& (s-6)^2 (s+t-6)^2P_{-2,0}\;,\\
            \mathcal{D}_{1,0} ={}& -(t-6)^2 \left((s-6)^2P_{-2,-2}-(t-8)^2P_{0,-4}+(s+t-6)^2P_{0,-2}\right)\;,\\
            \mathcal{D}_{1,1} ={}& (s-6)^2 \left((s-8)^2P_{-4,0}-(t-6)^2P_{-2,-2}-(s+t-6)^2P_{-2,0}\right)\;,\\
            \mathcal{D}_{2,0} ={}& (s-6)^2 (t-6)^2P_{-2,-2}\;.
        \end{split}
    \end{align}
    Comparing this to \cref{eq:n4_d}, we notice that it has the same structure and, in particular, the same shift operators as we had for $k_i=2$. We chose the Mellin ambiguity to reproduce the position space result, and we find that crossing is the same as for $k_i=2$, i.e. \cref{eq:crossing_reduced} with $\tilde u=u-4$.

    \subsubsection{6d \texorpdfstring{$\mathcal{N}=(2,0)$}{N=(2,0)} theory}
    We now move on to the 6d $\mathcal{N}=(2,0)$ theory. This theory is more complex, due to the presence of dependence on the $\sigma,\tau$ powers of the function on which we operator acts, as we can see from the position space solution, where this dependence appears in \cref{eq:6d_xyz}. This means that we must separate the reduced amplitude into components, as in \cref{eq:6d_red_amp_sep}, and have a $\mathcal{D}$ operator for each one. To simplify notation, we will define $\mathcal{D}^{lmn}$ as the operator that acts on a component $\widetilde{\mathcal{M}}_{k,lmn}(s,t)$ and write:
    \begin{equation}
        \mathcal{D}\widetilde{\mathcal{M}}(s,t;\sigma,\tau) = \sum_{l+m+n=k-2}\sigma^m\tau^n \mathcal{D}^{lmn}\widetilde{\mathcal{M}}_{k,lmn}(s,t)\;.
    \end{equation}
    Since there are now three operators, they will each carry their own Mellin ambiguity. This is because the Mellin ambiguities are essentially a redefinition of the space of function on which the operators act, and we can redefine each of the components of $\widetilde{\mathcal{M}}_{k,lmn}$ individually. We decompose each operator in a way analogous to \cref{eq:D_separation}:
    \begin{equation}
        \mathcal{D}^{lmn} = \sum_{i=0}^2\sum_{j=0}^{2-i}\sigma^i\tau^j \mathcal{D}_{i,j}^{lmn}\;.
    \end{equation}
    Then, we can write the solutions, starting with the $\mathcal{D}^{100}$ operator:
    \begin{align}
        \begin{split}
            \mathcal{D}_{0,0}^{100}={}& (s-12)^2 (t-12)^2 (t-10) (s+t-14) (s+t-12)^2P_{-2,-2}\\
            &+(s-6) (t-14)^2 (t-12)^2(s+t-14) (s+t-12)^2P_{0,-4}\\
            &-(s-6)(t-12)^2 (t-10) (s+t-10)^2 (s+t-12)^2P_{0,-2}\;,
        \end{split}
    \end{align}
    \begin{align}
        \begin{split}
            \mathcal{D}_{0,1}^{100}={}& -(s-14)^2 (s-12)^2 (t-10) (s+t-14) (s+t-12)^2P_{-4,0}\\
            &-(s-12)^2 (t-12)^2 (s+t-16)(s+t-14) (s+t-12)^2P_{-2,-2}\\
            &+(s-12)^2(t-10) (s+t-10)^2 (2 s+t-20) (s+t-12)^2P_{-2,0}\\
            &-(s-6) (t-14)^2 (t-12)^2 (s+t-14)(s+t-12)^2P_{0,-4}\\
            &+(s-6) (t-12)^2(s+t-10)^2 (s+2 t-24) (s+t-12)^2P_{0,-2}\\
            &-(s-6) (t-10) (s+t-10)^2 (s+t-8)^2(s+t-12)^2P_{0,0}\;,
        \end{split}
    \end{align}
    \begin{align}
        \begin{split}
            \mathcal{D}_{0,2}^{100}={}&(s-14)^2 (s-12)^2 (t-10) (s+t-14) (s+t-12)^2P_{-4,0}\\
            &+(s-12)^2 (s-6) (t-12)^2 (s+t-14)(s+t-12)^2P_{-2,-2}\\
            &-(s-12)^2 (s-6)(t-10) (s+t-10)^2 (s+t-12)^2P_{-2,0}\;,
        \end{split}
    \end{align}
    \begin{align}
        \begin{split}
            \mathcal{D}^{100}_{1,0}={}&-(s-14)^2 (s-12)^2 (t-10) (t-12)^2 (s+t-14)P_{-4,-2}\\
            &-(s-12)^2 (t-14)^2 (t-12)^2(s-t+4) (s+t-14)P_{-2,-4}\\
            &-(s-12)^2(t-10) (t-8) (t-12)^2 (s+t-12)^2P_{-2,-2}\\
            &+(s-6) (t-16)^2 (t-14)^2(t-12)^2 (s+t-14)P_{0,-6}\\
            &-(s-6) (t-14)^2(t-12)^2 (s+t-12)^2 (s+2 t-24)P_{0,-4}\\
            &+(s-6) (t-10) (t-12)^2 (s+t-12)^2(s+t-10)^2P_{0,-2}\;,
        \end{split}
    \end{align}
    \begin{align}
        \begin{split}
            \mathcal{D}^{100}_{1,1}={}&(s-16)^2 (s-14)^2 (s-12)^2 (t-10) (s+t-14)P_{-6,0}\\
            &+(s-14)^2 (s-12)^2 (t-12)^2(s-t+4) (s+t-14)P_{-4,-2}\\
            &-(s-14)^2(s-12)^2 (t-10) (s+t-12)^2 (2 s+t-20)P_{-4,0}\\
            &-(s-6) (s-12)^2 (t-14)^2 (t-12)^2(s+t-14)P_{-2,-4}\\
            &-(s-6) (s-4) (s-12)^2(t-12)^2 (s+t-12)^2P_{-2,-2}\\
            &+(s-6)(s-12)^2 (t-10) (s+t-12)^2 (s+t-10)^2P_{-2,0}\;,
        \end{split}
    \end{align}
    \begin{align}
        \begin{split}
            \mathcal{D}^{100}_{2,0}={}&(s-14)^2 (s-12)^2 (t-10) (t-12)^2 (s+t-14)P_{-4,-2}\\
            &+(s-12)^2 (s-6) (t-14)^2(t-12)^2 (s+t-14)P_{-2,-4}\\
            &-(s-12)^2(s-6) (t-10) (t-12)^2 (s+t-12)^2P_{-2,-2}\;.
        \end{split}
    \end{align}
    Then, the $\mathcal{D}^{010}$ operator:
    \begin{align}
        \begin{split}
            \mathcal{D}^{010}_{0,0}={}&(s-12)^2 (t-12)^2 (t-10) (s+t-18) (s+t-12)^2P_{-2,-2}\\
            &+(s-10) (t-14)^2 (t-12)^2(s+t-18) (s+t-12)^2P_{0,-4}\\
            &-(s-10)(t-12)^2 (t-10) (s+t-10)^2 (s+t-12)^2P_{0,-2}\;,
        \end{split}
    \end{align}
    \begin{align}
        \begin{split}
            \mathcal{D}^{010}_{0,1}={}&-(s-14)^2 (s-12)^2 (t-10) (s+t-18) (s+t-12)^2P_{-4,0}\\
            &-(s-12)^2 (t-12)^2 (s+t-20)(s+t-18) (s+t-12)^2P_{-2,-2}\\
            &+(s-12)^2(t-10) (s+t-10)^2 (2 s+t-28) (s+t-12)^2P_{-2,0}\\
            &-(s-10) (t-14)^2 (t-12)^2 (s+t-18)(s+t-12)^2P_{0,-4}\\
            &+(s-10) (t-12)^2(s+t-10)^2 (s+2 t-28) (s+t-12)^2P_{0,-2}\\
            &-(s-10) (t-10) (s+t-10)^2(s+t-8)^2 (s+t-12)^2P_{0,0}\;,
        \end{split}
    \end{align}
    \begin{align}
        \begin{split}
            \mathcal{D}^{010}_{0,2}={}&(s-14)^2 (s-12)^2 (t-10) (s+t-18) (s+t-12)^2P_{-4,0}\\
            &+(s-12)^2 (s-10) (t-12)^2 (s+t-18)(s+t-12)^2P_{-2,-2}\\
            &-(s-12)^2 (s-10)(t-10) (s+t-10)^2 (s+t-12)^2P_{-2,0}\;,
        \end{split}
    \end{align}
    \begin{align}
        \begin{split}
            \mathcal{D}^{010}_{1,0}={}&-(s-14)^2 (s-12)^2 (t-10) (t-12)^2 (s+t-18)P_{-4,-2}\\
            &-(s-12)^2 (t-14)^2 (t-12)^2(s-t) (s+t-18)P_{-2,-4}\\
            &-(s-12)^2 (t-10)(t-8) (t-12)^2 (s+t-12)^2P_{-2,-2}\\
            &+(s-10) (t-16)^2 (t-14)^2(t-12)^2 (s+t-18)P_{0,-6}\\
            &-(s-10) (t-14)^2(t-12)^2 (s+t-12)^2 (s+2 t-28)P_{0,-4}\\
            &+(s-10) (t-10) (t-12)^2 (s+t-12)^2(s+t-10)^2P_{0,-2}\;,
        \end{split}
    \end{align}
    \begin{align}
        \begin{split}
            \mathcal{D}^{010}_{1,1}={}&(s-16)^2 (s-14)^2 (s-12)^2 (t-10) (s+t-18)P_{-6,0}\\
            &+(s-14)^2 (s-12)^2 (t-12)^2 (s-t)(s+t-18)P_{-4,-2}\\
            &-(s-14)^2 (s-12)^2(t-10) (s+t-12)^2 (2 s+t-28)P_{-4,0}\\
            &-(s-10) (s-12)^2 (t-14)^2 (t-12)^2(s+t-18)P_{-2,-4}\\
            &-(s-10) (s-8) (s-12)^2(t-12)^2 (s+t-12)^2P_{-2,-2}\\
            &+(s-10)(s-12)^2 (t-10) (s+t-12)^2 (s+t-10)^2P_{-2,0}\;,
        \end{split}
    \end{align}
    \begin{align}
        \begin{split}
            \mathcal{D}^{010}_{2,0}={}&(s-14)^2 (s-12)^2 (t-10) (t-12)^2 (s+t-18)P_{-4,-2}\\
            &+(s-12)^2 (s-10) (t-14)^2(t-12)^2 (s+t-18)P_{-2,-4}\\
            &-(s-12)^2(s-10) (t-10) (t-12)^2 (s+t-12)^2P_{-2,-2}\;.
        \end{split}
    \end{align}
    Finally, the $\mathcal{D}^{001}$ operator:
    \begin{align}
        \begin{split}
            \mathcal{D}^{001}_{0,0}={}&(s-12)^2 (t-12)^2 (t-6) (s+t-14) (s+t-12)^2P_{-2,-2}\\
            &+(s-10) (t-14)^2 (t-12)^2(s+t-14) (s+t-12)^2P_{0,-4}\\
            &-(s-10)(t-12)^2 (t-6) (s+t-10)^2 (s+t-12)^2P_{0,-2}\;,
        \end{split}
    \end{align}
    \begin{align}
        \begin{split}
            \mathcal{D}^{001}_{0,1}={}&-(s-14)^2 (s-12)^2 (t-6) (s+t-14) (s+t-12)^2P_{-4,0}\\
            &-(s-12)^2 (t-12)^2 (s+t-16)(s+t-14) (s+t-12)^2P_{-2,-2}\\
            &+(s-12)^2(t-6) (s+t-10)^2 (2 s+t-24) (s+t-12)^2P_{-2,0}\\
            &-(s-10) (t-14)^2 (t-12)^2 (s+t-14)(s+t-12)^2P_{0,-4}\\
            &+(s-10) (t-12)^2(s+t-10)^2 (s+2 t-20) (s+t-12)^2P_{0,-2}\\
            &-(s-10) (t-6) (s+t-10)^2 (s+t-8)^2(s+t-12)^2P_{0,0}\;,
        \end{split}
    \end{align}
    \begin{align}
        \begin{split}
            \mathcal{D}^{001}_{0,2}={}&(s-14)^2 (s-12)^2 (t-6) (s+t-14) (s+t-12)^2P_{-4,0}\\
            &+(s-12)^2 (s-10) (t-12)^2 (s+t-14)(s+t-12)^2P_{-2,-2}\\
            &-(s-12)^2 (s-10)(t-6) (s+t-10)^2 (s+t-12)^2P_{-2,0}\;,
        \end{split}
    \end{align}
    \begin{align}
        \begin{split}
            \mathcal{D}_{1,0}^{001}={}&-(s-14)^2 (s-12)^2 (t-6) (t-12)^2 (s+t-14)P_{-4,-2}\\
            &-(s-12)^2 (t-14)^2 (t-12)^2(s-t-4) (s+t-14)P_{-2,-4}\\
            &-(s-12)^2 (t-6)(t-4) (t-12)^2 (s+t-12)^2P_{-2,-2}\\
            &+(s-10) (t-16)^2 (t-14)^2(t-12)^2 (s+t-14)P_{0,-6}\\
            &-(s-10) (t-14)^2(t-12)^2 (s+t-12)^2 (s+2 t-20)P_{0,-4}\\
            &+(s-10) (t-6) (t-12)^2 (s+t-12)^2(s+t-10)^2P_{0,-2}\;,
        \end{split}
    \end{align}
    \begin{align}
        \begin{split}
            \mathcal{D}^{001}_{1,1}={}&(s-16)^2 (s-14)^2 (s-12)^2 (t-6) (s+t-14)P_{-6,0}\\
            &+(s-14)^2 (s-12)^2 (t-12)^2(s-t-4) (s+t-14)P_{-4,-2}\\
            &-(s-14)^2(s-12)^2 (t-6) (s+t-12)^2 (2 s+t-24)P_{-4,0}\\
            &-(s-10) (s-12)^2 (t-14)^2 (t-12)^2(s+t-14)P_{-2,-4}\\
            &-(s-10) (s-8) (s-12)^2(t-12)^2 (s+t-12)^2P_{-2,-2}\\
            &+(s-10)(s-12)^2 (t-6) (s+t-12)^2 (s+t-10)^2P_{-2,0}\;,
        \end{split}
    \end{align}
    \begin{align}
        \begin{split}
            \mathcal{D}^{001}_{2,0}={}&(s-14)^2 (s-12)^2 (t-6) (t-12)^2 (s+t-14)P_{-4,-2}\\
            &+(s-12)^2 (s-10) (t-14)^2(t-12)^2 (s+t-14)P_{-2,-4}\\
            &-(s-12)^2(s-10) (t-6) (t-12)^2 (s+t-12)^2P_{-2,-2}\;.
        \end{split}
    \end{align}
    An important thing to note about these is that all three operators share the same shift operator content, differing only in the coefficients. The Mellin ambiguities were chosen to match the solution from position space. This leads to the same crossing equation as we had for $k_i=2$, i.e. \cref{eq:crossing_reduced} with $\tilde u = u-6$.

    It seems likely that, in the other theories, and at higher $k$ in this one, we would have to similarly consider separate operators for each component of the reduced amplitude. We will not consider here $k_i=3$ in the other theories for the sake of brevity.

    \subsection{A mixed correlator: \texorpdfstring{$\expval{kk22}$}{<kk22>}}\label{sec:22kk}
    In this section, we will consider a mixed correlator, one of operators with $k_i$ values $k_1=k_2\equiv k$ and $k_3=k_4=2$. We will focus on the case of 4d $\mathcal{N}=4$ SYM, since the solution is known in this case from position space. Our objective is to show that the method of \cref{sec:mellin_sol} works even in the case of unequal correlators. 

    In this case, the extremality is $L=2$, so the full amplitudes are polynomials of order 2 in $\sigma,\tau$. We can use the methods of \cref{sec:mellin_sol} without any modification, and as usual separate $\mathcal{D}$ into components as in \cref{eq:D_separation}. Then, the solution is:
    \begin{align}
        \begin{split}
        \mathcal{D}_{0,0} ={}& (k-t+2)^2 (-k+s+t-2)^2P_{0,-2}\;,\\
        \mathcal{D}_{0,1} ={}& (s-4) (2 k-s) (-k+s+t-2)^2 P_{-2,0}-(k-t+2)^2(-k+s+t-2)^2 P_{0,-2}\;,\\
        &+(-k+s+t)^2 (-k+s+t-2)^2P_{0,0}\;,\\
        \mathcal{D}_{0,2} ={}& -(s-4) (2 k-s) (-k+s+t-2)^2P_{-2,0}\;,\\
        \mathcal{D}_{1,0} ={}&(s-4) (2 k-s) (k-t+2)^2P_{-2,-2}+(k-t+4)^2 (k-t+2)^2P_{0,-4}\;,\\
        &-(k-t+2)^2 (-k+s+t-2)^2P_{0,-2}\;,\\
        \mathcal{D}_{1,1} ={}&(s-4)(2 k-s)(k-t+2)^2P_{-2,-2}+(s-6) (s-4) (2 k-s)(2 k-s+2)P_{-4,0}\;,\\
        &+(s-4) (2 k-s) (-k+s+t-2)^2P_{-2,0}\;,\\
        \mathcal{D}_{2,0}={}&-(s-4) (2 k-s) (k-t+2)^2P_{-2,-2}\;.
        \end{split}
    \end{align}
    and we can check that we recover \cref{eq:n4_d} if we set $k=2$ in the above. We can also check that this matches with the solution from position space.

    \bibliographystyle{JHEP}
    \bibliography{bibliography.bib}
\end{document}